\documentclass[usenatbib]{mn2e} 
\usepackage{epsfig}
\usepackage{amsmath}
\def\gsim{\mathrel{\raise0.35ex\hbox{$\scriptstyle >$}\kern-0.6em
\lower0.40ex\hbox{{$\scriptstyle \sim$}}}}
\def\lsim{\mathrel{\raise0.35ex\hbox{$\scriptstyle <$}\kern-0.6em
\lower0.40ex\hbox{{$\scriptstyle \sim$}}}}

\title[Red-sequence LFs from CFHTLS clusters]{Recent Arrival of Faint Cluster Galaxies on the Red-sequence: Luminosity Functions from 119 square degrees of CFHTLS}
\author[Lu et al.]{Ting Lu\thanks{E-mail: t5lu@sciborg.uwaterloo.ca}, David G. Gilbank, Michael L. Balogh, Adam Bognat\\
Department of Physics and Astronomy, University of Waterloo, Waterloo, Ontario,
N2L 3G1, Canada\\
}
\date{\today}

\begin{document}

\maketitle

\begin{abstract}
The global star formation rate has decreased significantly since $z\sim 1$, for reasons that are not well understood. Red-sequence galaxies, dominating in galaxy clusters, represent the population that have had their star formation shut off, and may therefore be the key to this problem. In this work, we select 127 rich galaxy clusters at $0.17\leq z \leq 0.36$, from 119 square degrees of the Canada-France-Hawaii Telescope Legacy Survey (CFHTLS) optical imaging data, and construct the $r'-$band red-sequence luminosity functions (LFs). We show that the faint end of the LF is very sensitive to how  red-sequence galaxies are selected, and an optimal way to minimise the contamination from the blue cloud is to mirror galaxies on the redder side of the colour-magnitude relation (CMR). The LFs of our sample have a significant inflexion centred at $M_{r'}\sim -18.5$, suggesting a mixture of two populations. Combining our survey with low redshift samples constructed from the Sloan Digital Sky Survey, we show that there is no strong evolution of the faint end of the LF (or the red-sequence dwarf-to-giant ratio) over the redshift range $0.2\lsim z \lsim 0.4$, but from $z\sim 0.2$ to $z\sim 0$ the relative number of red-sequence dwarf galaxies has increased by a factor of $\sim 3$, implying a significant build-up of the faint end of the cluster red-sequence over the last 2.5 Gyr.

\end{abstract}


\section{Introduction}
\renewcommand\thefootnote{\fnsymbol{footnote}}

It has been established that the galaxy population today can be broadly divided into two categories: those that have red colours, consisting of mostly non-star-forming galaxies (the `red-sequence'); and those with blue colours and active star formation (the `blue cloud'). This colour bimodality can be modelled by a sum of two Gaussian distributions \citep{balogh04b,balogh04,baldry}, and it has been observed out to $z\sim 1$ \citep{bell}. The red population exhibits a tight correlation between colour and magnitude, with brighter galaxies being redder. Due to the age-metallicity degeneracy \citep{worthey}, the slope of this colour-magnitude relation (CMR) of the red population could be attributed to either an age or metallicity sequence. The study by  \cite{met_mag} concluded that metallicity variation dominates the slope, because of its relatively slow evolution. Nonetheless, some age variation along the red-sequence is also detected \citep[e.g.][]{trager00,nelan,SH,steve}.

The origin of red-sequence galaxies is still an open question. 
Studies by \cite{bell} and \cite{faber07} concluded that brighter red-sequence galaxies are built up through dry mergers since $z\sim 1$, based on observations that the $B-$band luminosity density of red-sequence galaxies remains constant since $z\sim 0.9$. With the dimming of galaxies, the luminosity density would be overproduced if the number density of red-sequence galaxies remained constant, as in the pure passive evolution scenario. However, other studies by, for example, \cite{cimatti06} and \cite{scarlata}, showed that the number density of the massive red galaxies remains constant and their luminosity function is consistent with passive evolution. It is the number density of less-massive galaxies that decreases rapidly with increasing redshift, and this can be explained by a gradual quenching of star formation in those galaxies. Therefore, no significant amount of dry mergers are required to explain the formation of massive red-sequence galaxies. Which of the two proposed scenarios is more important is still not clear.

As to how the faint end of the red-sequence builds up, it is even less clear. Especially in galaxy clusters, different studies have yielded conflicting results. For example, observations by \cite{lucia}, \cite{stott} and \cite{gilbanklf} found a significant deficit of faint red-sequence galaxies in high redshift clusters, compared to low redshift clusters. This supports a scenario where low mass galaxies have their star formation shut off and move onto the red-sequence recently ($z\lsim 1$). However, \cite{andreon08} and  \cite{crawford} found no evolution of the faint-end slope spanning $0<z<1.3$, which implies the build-up of the faint red-sequence galaxies was completed by $z\sim 1.3$. 

 Current galaxy formation models  \citep[e.g.][]{bowermodel}  produce too many red galaxies in clusters and groups that are not observed \citep{wolf05,wm06,weinmann06,baldry06}, indicating a fundamental problem in our understanding of how star formation is quenched in dense environments. More precise observations are needed to resolve this problem.

In this work, we make use of optical imaging data from the Canada-France-Hawaii Telescope Legacy
Survey (CFHTLS) data to construct a large sample of 127 clusters and study their red-sequence luminosity functions over the redshift range  $0.17\leq z\leq 0.36$. In \S \ref{data} and \S \ref{detect} we describe  the data, and the cluster detection algorithm. The properties of our cluster catalogue, and a local comparison sample are described in \S \ref{prop} and \S \ref{local}. We present our methods for measuring the red-sequence luminosity function and dwarf-to-giant ratio in \S \ref{rel}, and the results in \S \ref{results}. In \S \ref{disc} we discuss the comparison with literature and possible systematics, and conclude in \S \ref{con}.

Throughout this paper, we assume a cosmology 
with $\Omega_m=0.3$, $\Omega_{\Lambda}=0.7$ and
$H_o$=70 km s$^{-1}$ Mpc$^{-1}$. All magnitudes are in the AB system unless otherwise specified.

\section{Data}\label{data}
\subsection{The Survey}
The CFHTLS is a joint Canadian and French imaging survey in  $u^*$, $g'$, $r'$, $i'$ and $z'$ filters using MegaCam, with an approximately 1x1 square degree field of view. We use the ``Wide'' and ``Deep'' surveys in this work. The Wide survey covers a total area of 171 square degrees with a total exposure time of about 2500 seconds in $g'$ band, 1000 seconds in $r'$ band and 4300 seconds in $i'$ band per pointing.  The Deep survey contains four 1 square degree fields with exposure times ranging from 33 hours in $u^*$ band to 132 hours in $i'$ band. 

The survey started in 2003 and is now complete, but not all data have been processed and released yet. The data we used in this work are from data release T0004 (released internally July 3, 2007). We carry out our cluster detection using the Wide survey data over a total area of 119 square degrees (details summarized in Table \ref{ar}), where the photometry in all three of the filters, $g'$, $r'$ and $i'$, is available. We use data from one of the four Deep fields, D1, as auxiliary data to examine surface brightness selection effects (\S \ref{sbeff})  and other possible systematics (\S \ref{d1}).

\begin{table}
\caption{Total area covered by the Wide survey and area used in this work.}
\begin{tabular}{|cc|cc|cc|c|c|} \hline
$ $ & W1 & W2 & W3 & W4 \\
\hline
Total  (sq. deg) & 72 & 25 & 49 & 25 \\
\hline
Used in this work (sq. deg) & 42 & 20 & 41 & 16  \\
\hline
\end{tabular}\label{ar}
\end{table}

\addtocounter{footnote}{-1}
\subsection{Photometry}\label{phot}
For our analysis, we are interested in measuring the colours and total magnitudes of galaxies. The photometric catalogue we use was produced by TERAPIX (Traitement Elementaire, Reduction et Analyse des PIXels de megacam). Magnitudes in the catalogue are measured using SExtractor \citep{sextractor} within different apertures. We take $mag\_{auto}$, where flux is measured within the Kron radius, as the total magnitude of galaxies. 

We measure colours of galaxies using magnitudes within a fixed aperture of 4.7 arcsec in diameter. The seeing$^\star$\footnote{$^\star$defined
by TERAPIX as twice the median half-light radius of a selection of
point sources on each CCD as measured by SExtractor.} of all images in this release is better than 1.3 arcsec, and the typical seeing of each pointing is $\sim$0.94 arcsec in $g'$, $\sim$0.86 arcsec in $r'$ and $\sim$0.81 arcsec in $i'$. For each pointing, the maximum difference in the seeing of the stacked, single-filter images between $g'$, $r'$ and $i'$ band is less than 0.5 arcsec in the worst case $-$ small compared with the radius of the aperture used to measure the colour. Therefore, we do not convolve the images in different filters to the same seeing. TERAPIX compared the photometry of objects in the overlap regions between the CFHTLS and the Sloan Digital Sky Survey (SDSS) \citep{sloan}, and found that the mean offset in $g'$, $r'$,and $i'$ filters in each individual pointing ranges from $\sim 0.01$ to $\sim 0.05$ mag.

The galactic E(B-V) extinction calculated using the \cite{ext} dust map is provided in the catalogue for each object. We use
the A/E(B-V) values given in \cite{ext} in SDSS \citep{sloan}  $u$, $g$, $r$, $i$, and $z$ 
filters to calculate the extinction correction and apply this to
the observed magnitude. The median extinction in $r'$ band is 0.072 mag in W1, 0.056 in W2, 0.032 in W3 and 0.217 in W4.

\subsubsection{Depth and Photometric Uncertainty}\label{sbeff}

\begin{figure}
\leavevmode \epsfysize=8cm
\includegraphics[width=45mm]{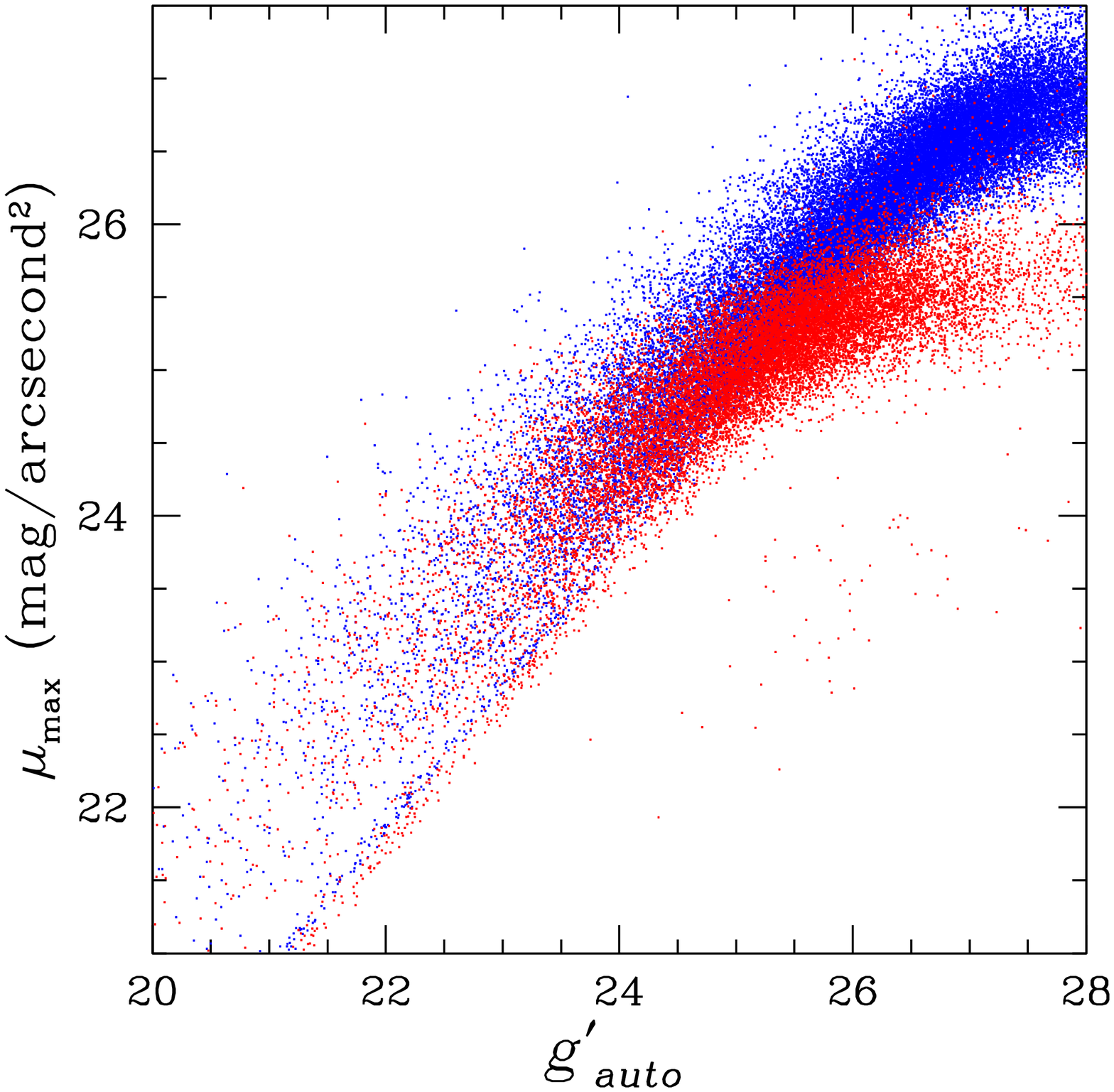}\includegraphics[width=45mm]{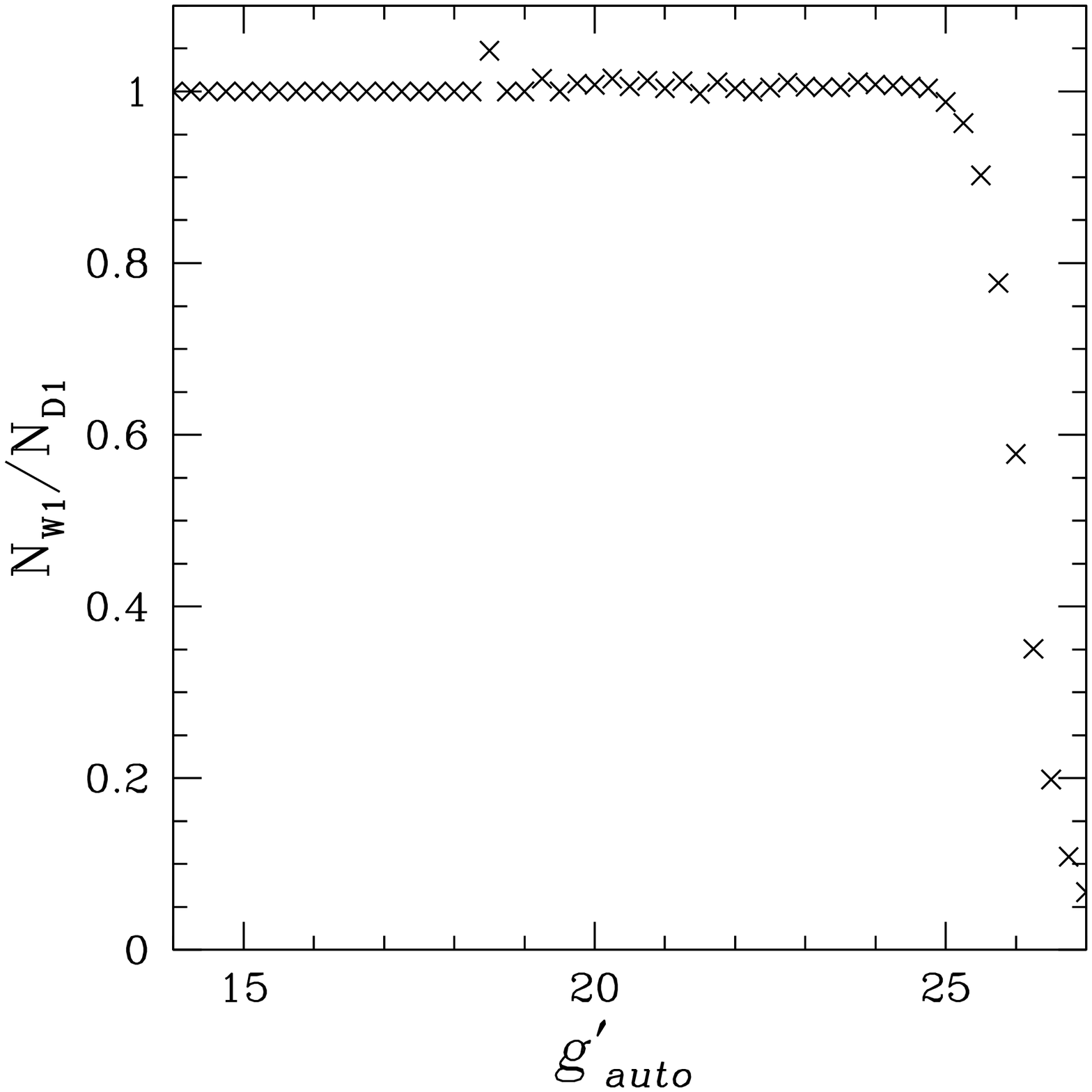}
\includegraphics[width=45mm]{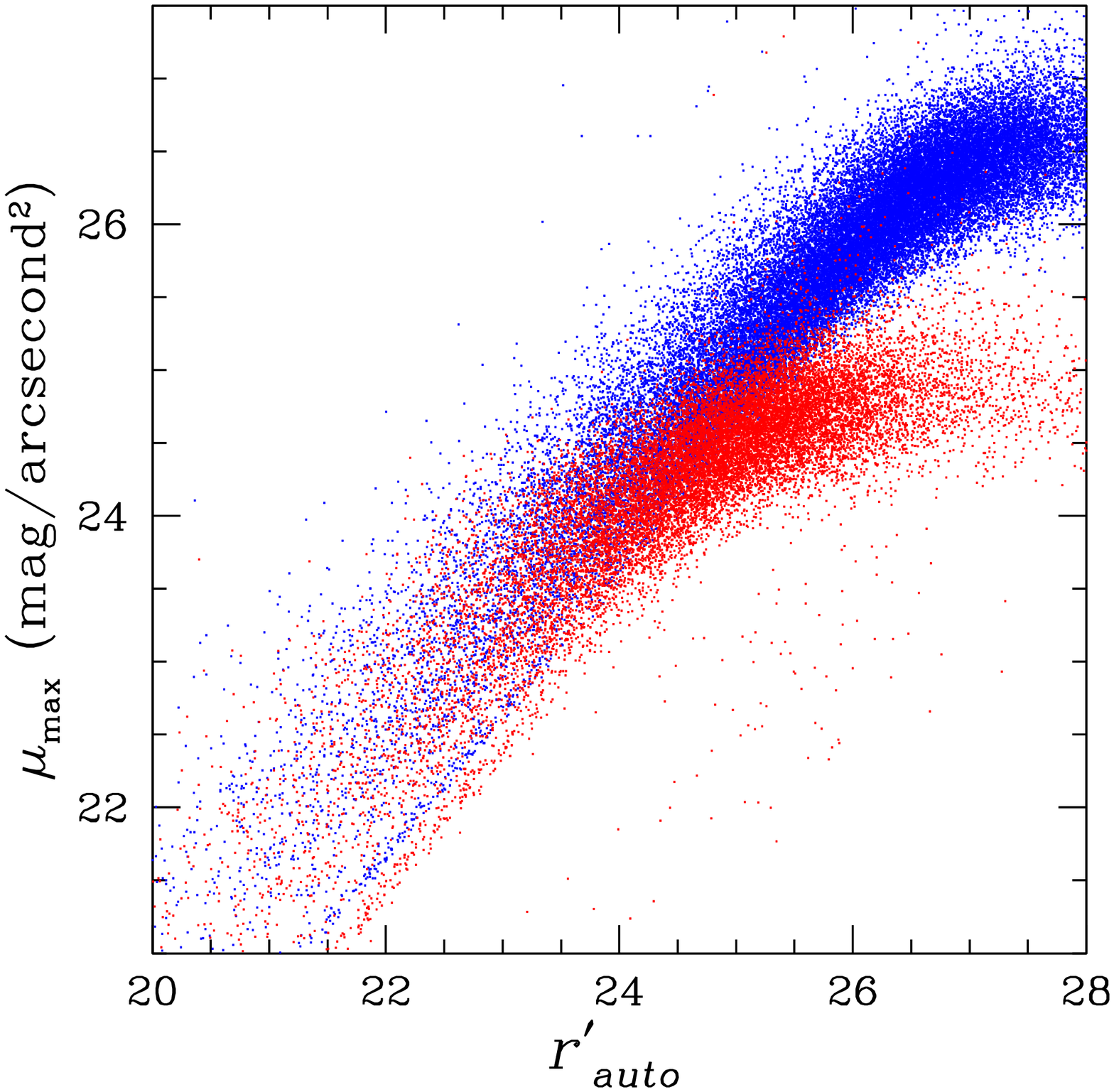}\includegraphics[width=45mm]{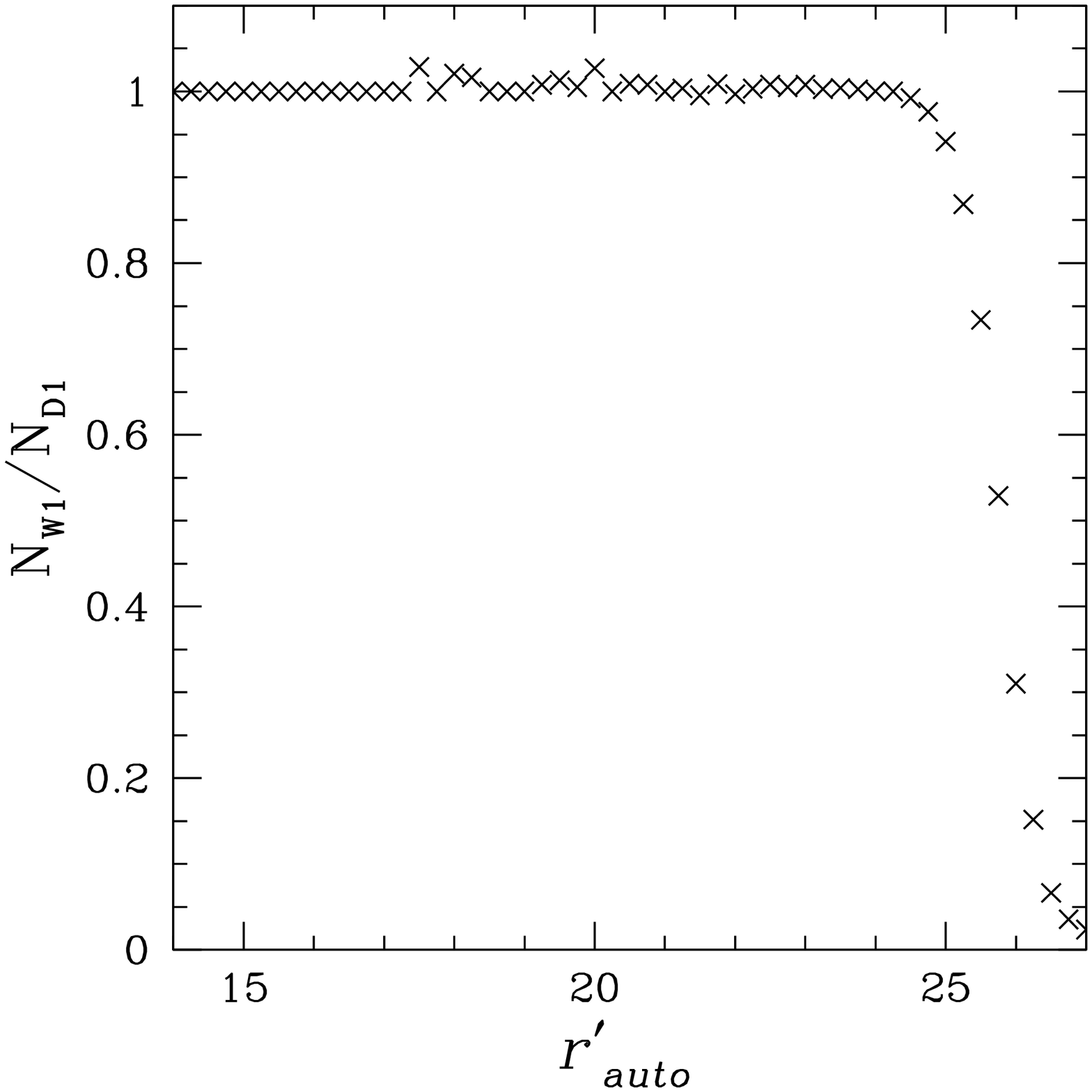}
 \caption{Left column: central surface brightness, $\mu_{max}$, as a function of $g'$ (top) and $r'$ (bottom) magnitude. Red points are objects in W1 and blue points D1. Right column: fraction of sources in D1 that are also detected in W1, as a function of magnitude. }\label{sb}
\end{figure}

To examine how surface brightness might affect the completeness of the sample, we plot the surface brightness for the brightest pixel, $\mu_{max}$ (which is a good proxy for the central surface brightness, measured using SExtractor), vs. magnitude for galaxies that are in the overlap region between D1 and W1, as shown in the left column in Figure \ref{sb}. Blue points represent objects in D1 and red points W1. In $g'$ band, at about $g'\sim25$, W1 starts to become incomplete due to the surface brightness detection limit. To quantify this, we calculate the fraction of sources in D1 that are also detected in W1 as a function of magnitude. As shown in the right column of Figure \ref{sb}, W1 is $\sim$ 100 per cent complete at $g'=25$, and $r'=24$. Since the $(g'-r')$ colour of red galaxies at the highest redshift we focus on in this paper is around 1.8 (Figure \ref{CMgr}), we make a cut at $r'=23.2$ to ensure completeness in the $g'$ band.

\begin{figure}
\leavevmode \epsfysize=8cm \epsfbox{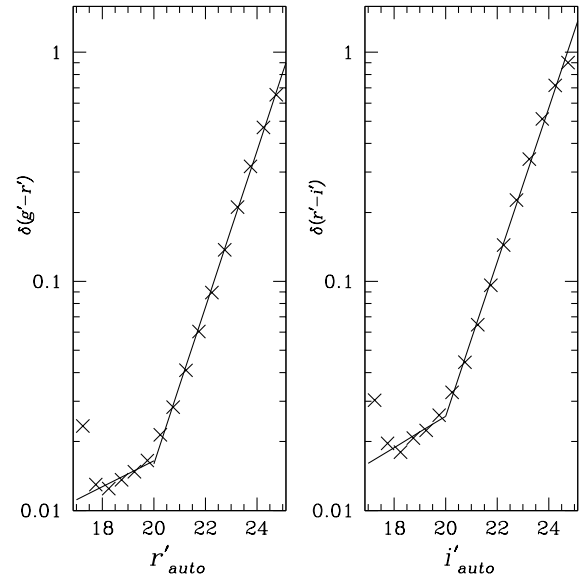} \caption{Left panel: the standard deviation of $(g'-r')$ colour differences in overlapping pointings as a function of $r'$ magnitude, fitted by two straight lines on logarithmic scale. Right panel: the standard deviation of  $(r'-i')$ colour differences as a function of $i'$ magnitude. \label{grerrlinfit}}
\end{figure}

\addtocounter{footnote}{-1}
To determine the uncertainty on the colour, we examine the difference in the colours measured from each pointing, for galaxies in the overlap regions between pointings.$^\ddagger$\footnote{$^\ddagger$The residual mean offset in colour between two overlapping pointings is $<0.05$ mag, and we correct for this residual offset before we stack all the overlap regions together to examine the distribution of the difference between colours measured from two overlapping pointings.} In Figure \ref{grerrlinfit}, we plot the standard deviation of $(g'-r')$ (left panel) and $(r'-i')$ (right panel) colour differences as a function of magnitude, and fit the logarithm of it by two straight lines. At the magnitude limit of $r'=23.2$ of our catalogue, the uncertainty on the $(g'-r')$ colour is  about 0.2 mag, which makes isolating red-sequence galaxies from the contamination of the blue cloud less reliable at fainter magnitudes (see \S \ref{rel}).

\subsection{Star-Galaxy Separation}

\begin{figure}
\leavevmode \epsfysize=8cm
\epsfbox{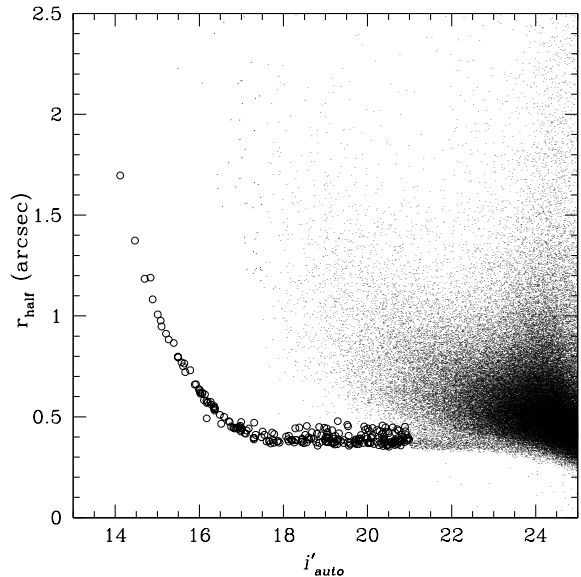}
\caption{Half-light radius vs. magnitude. Points represent a subset of the galaxies that are in our final galaxy catalogue, and circles represent stars. See text for details on star-galaxy separation.\label{sgsep}}
\end{figure}

TERAPIX classified objects in the magnitude range $17.5<i'<21.0$, using the stellar locus in the half-light radius vs. magnitude plot, in a series of $10$ arcmin cells distributed over each MegaCam stack. All objects fainter than $i'=21.0$ are considered galaxies. At magnitudes brighter than $i'=17.5$, we keep objects that have SExtractor parameter $class\_star\leq 0.98$, but exclude those on the stellar locus, as galaxies. Objects in our final galaxy catalogue are plotted as small points in Figure \ref{sgsep}, and stars are represented by circles. Only a subset of the data is plotted for clarity.

\section{Cluster Detection}\label{detect}
The cluster detection method we use is based on the spirit of the cluster-red-sequence (CRS) method \citep{GY00}, but with some modifications. The foundation of the CRS technique is that in every cluster there is a population of red galaxies that forms a tight red-sequence in the colour-magnitude diagram (CMD), which can be used as an overdensity indicator. Four major steps are involved in the cluster detection: 1) model the CMD at different redshifts; 2) select a subsample of galaxies that belongs to each colour slice; 3) count galaxies around a grid of positions in the sky and estimate the background in the same way; 4) select overdensities as cluster candidates.  Below, we describe each step in detail.

\subsection{Model CMD}\label{mod}
First we model the CMD for the red-sequence galaxies. Since we only need a model that produces the correct CMR over a small redshift range, and we will calibrate it empirically, a simple model is sufficient. We use a single burst model from  \cite{galexev} with a Salpeter IMF and formation redshift of $z_f\sim$2. We follow the passive evolution of galaxies with different metallicities, and calibrate the model by defining a magnitude-metallicity relation so that it reproduces the observed CMD of E+S0 galaxies in Coma (slope $-0.0743$, zero point 4.21, \citealt{bower92}).  The resulting magnitude-metallicity relation is in good agreement with the result from a more complicated model by \cite{met_mag}.

Under the assumption that this magnitude-metallicity relation does not evolve with redshift, we can define for each metallicity a correlation between the model colour and magnitude for the passively-evolving galaxies at each redshift. We fit the resulting CMR with a straight line. To model $m^*$ at different redshifts, we take the $m^*$ at $z=0$ estimated by  \cite{Blanton01} using SDSS data, transform it into CFHTLS filters, and evolve it using the same model with solar metallicity. 

\begin{figure}

\leavevmode \epsfysize=8cm
\epsfbox{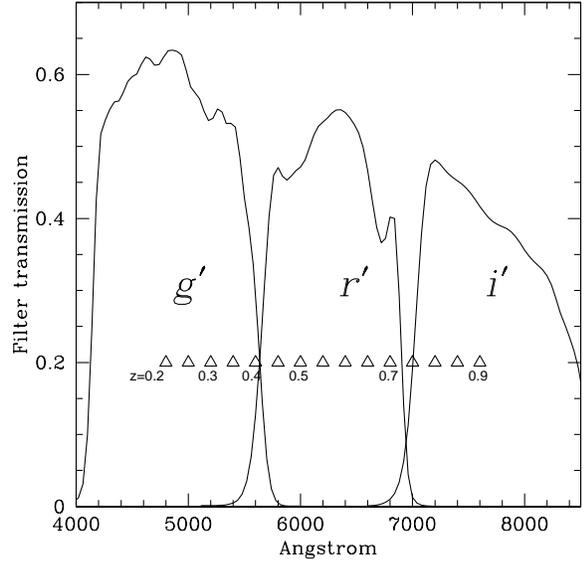}
\caption{Position of the 4000\AA \ break at different redshifts in CFHTLS filters. Solid curves are total filter transmissions (filter+mirror+optics+CCD), and triangles indicate the positions of the 4000\AA \ break at different redshifts. \label{4000A}}
\end{figure}

\addtocounter{footnote}{-1}
Figure \ref{4000A} shows the CFHTLS filter transmission curves$^\dagger$\footnote{$^\dagger$http://www3.cadc-ccda.hia-iha.nrc-cnrc.gc.ca/megapipe/docs\\/filters.html}, and into which filters the 4000\AA \ break (the most prominent feature in a red-sequence galaxy's spectrum) falls at different redshifts. When the 4000\AA \ break is in the $g'$ filter, the $(g'-r')$ colour is very sensitive to a small shift of the break position, and thus gives the best redshift resolution. When the break is approaching  the boundary between $g'$ and $r'$ filters, the $(g'-r')$ colour becomes degenerate. Therefore, we limit our study in this work to $z<0.36$.

\begin{figure}
\leavevmode \epsfysize=8cm \epsfbox{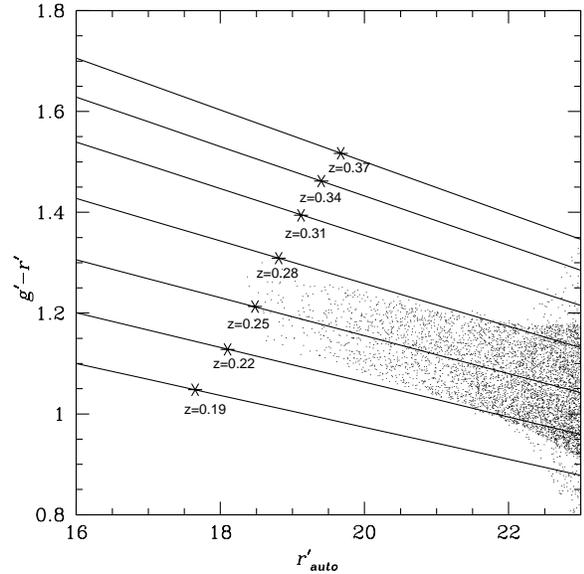}
\caption{Model CMD in CFHTLS filters as a function of redshift  with $m^*$ indicated. The redshift interval is $\triangle z=0.03$. The $z\sim 0$ relation is calibrated to Coma \citep{bower92}. The CMD is evolved passively, with a fixed magnitude-metallicity relation. Points represent a subset of galaxies  from W1 that are associated with the colour slice at $z=0.25$ (see \S \ref{sub}).  \label{CMgr}}
\end{figure}

Figure \ref{CMgr}  shows the modelled $(g'-r')$
vs. $r'$ diagram as a function of redshift, with
$m^*$ indicated. The redshift interval between slices is 0.03.  The $m^*$ and colour at $m^*$ are tabulated in Table \ref{tabmstar}, with a small, empirical correction applied to the redshifts as discussed in \S \ref{ppz}.

\begin{table}
\begin{center}
\caption{Model $m^*$ and colour at $m^*$ as a function of redshift.}
\begin{tabular}{|cc|cc|cc|c|c|} \hline
$z$ & $m^*_{r'}$ & $(g'-r')$\ at $m^*_{r'}$  \\
\hline
        0.19  &      17.65    &    1.048\\
\hline
        0.22   &      18.1    &    1.128\\
\hline
        0.25  &      18.48   &     1.213\\
\hline
        0.28  &      18.81   &     1.309\\
\hline
        0.31  &      19.12   &     1.394\\
\hline
        0.34  &       19.4    &    1.462\\
\hline
        0.37   &     19.67    &    1.517\\
\hline
\end{tabular}\label{tabmstar}
\end{center}
\end{table}

\subsection{Subsamples in each Colour Slice}\label{sub}
The next step is to construct a subsample of galaxies for each redshift based on the colours of the galaxies. At each redshift, we define a colour slice centred on the model CMR, with the half width determined  by the intrinsic  scatter in the colours of red-sequence galaxies, 0.075 \citep{GY00}. The colour slices overlap with each other to avoid missing clusters on the edge of each colour slice. We then calculate, for each galaxy, the probability that it belongs to a specific colour slice, assuming that the probability distribution of the colour of each galaxy is Gaussian, with $\sigma=\triangle c$, dominated by the colour errors estimated from the overlapping regions as shown in Figure \ref{grerrlinfit}.  For each colour slice, all
galaxies with probabilities larger than $10$ per cent are selected.
A subset of the galaxies from W1 that are
associated with the colour slice at $z=0.25$ are shown as points in Figure \ref{CMgr}. At the faint end, due to the larger colour uncertainties, galaxies
belonging to that colour slice spread out of the boundary of the
slice. 

This colour-weighting step is the same as prescribed by \cite{GY00}. In the original CRS technique, after the colour weighting, a magnitude weighting is applied to downweight the fainter galaxies because the contrast between cluster and the field is lower at faint magnitudes. We do not want to bias our detected clusters to a certain luminosity function shape, so we do not apply any magnitude-weighting here. However, to avoid high contamination from the field at the very faint magnitudes, we only use galaxies brighter than $m^*+2$ for detection. For the purpose of our study here, we only need a sample of rich clusters without worrying about the completeness of poorer systems, which further justifies the modification we make here.

\subsection{Significance Map and Detection}
In each colour slice, we count the number of galaxies in the subsample within a circle of radius $\sim0.5$ Mpc (the typical size of a cluster core) around a grid of positions $\sim0.1$ Mpc apart in the sky. This gives the cluster+field count at each position. The field count is estimated in the same way, but from the average of 500 random positions in the sky. The difference between the total number count
and the background count at each grid position divided by the
rms deviation of the background distribution of the 500 random positions (comparable to Poisson uncertainty), $\sigma_f$, indicates the significance of
the overdensity smoothed on a $0.5$ Mpc scale at that position, i.e.
\begin{equation}
\sigma ={N_{cluster+field}-N_{field}\over {\sigma_f}}.
\end{equation}

 FITS images of the significance
maps for each colour slice are created. SExtractor  is then run on
those images to pick out peaks on the significance maps. We keep everything that has at least 1 pixel above $5\sigma$ as cluster candidates.

We now have a list of crude positions and significance of cluster
candidates. If multiple peaks are detected within $\sim$1 Mpc from each other, either in the same colour slice or adjacent slices, only the one that has the highest significance will be kept. In the following sections, we refine the cluster catalogue, by using additional colour information, and improving the centring and redshift resolution.

\begin{figure}
\leavevmode \epsfysize=8cm \epsfbox{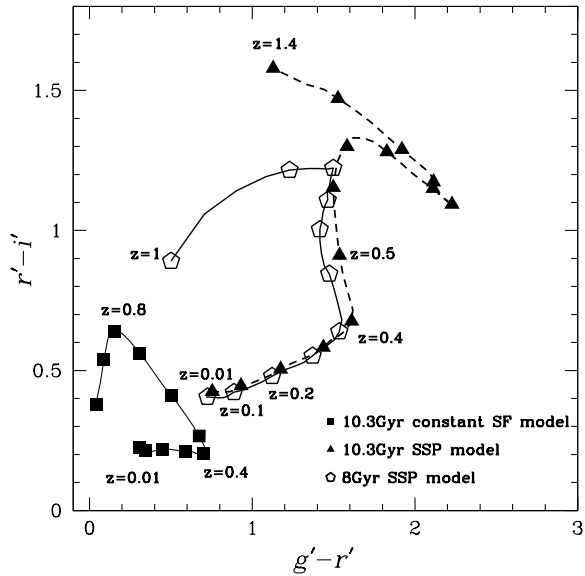}
 \caption{Evolutionary tracks of $(g'-r')$ vs $(r'-i')$ with redshift for three different populations produced by the model of \protect\cite{galexev}. Squares are a population that formed 10.3 Gyr ago and have constant star formation. Pentagons represents a 8 Gyr old single-burst model and triangles a 10.3 Gyr single-burst model. The lines are to guide the eyes.\label{grritrack}}
\end{figure}

\begin{figure*}
\epsfysize=18cm
\epsfbox{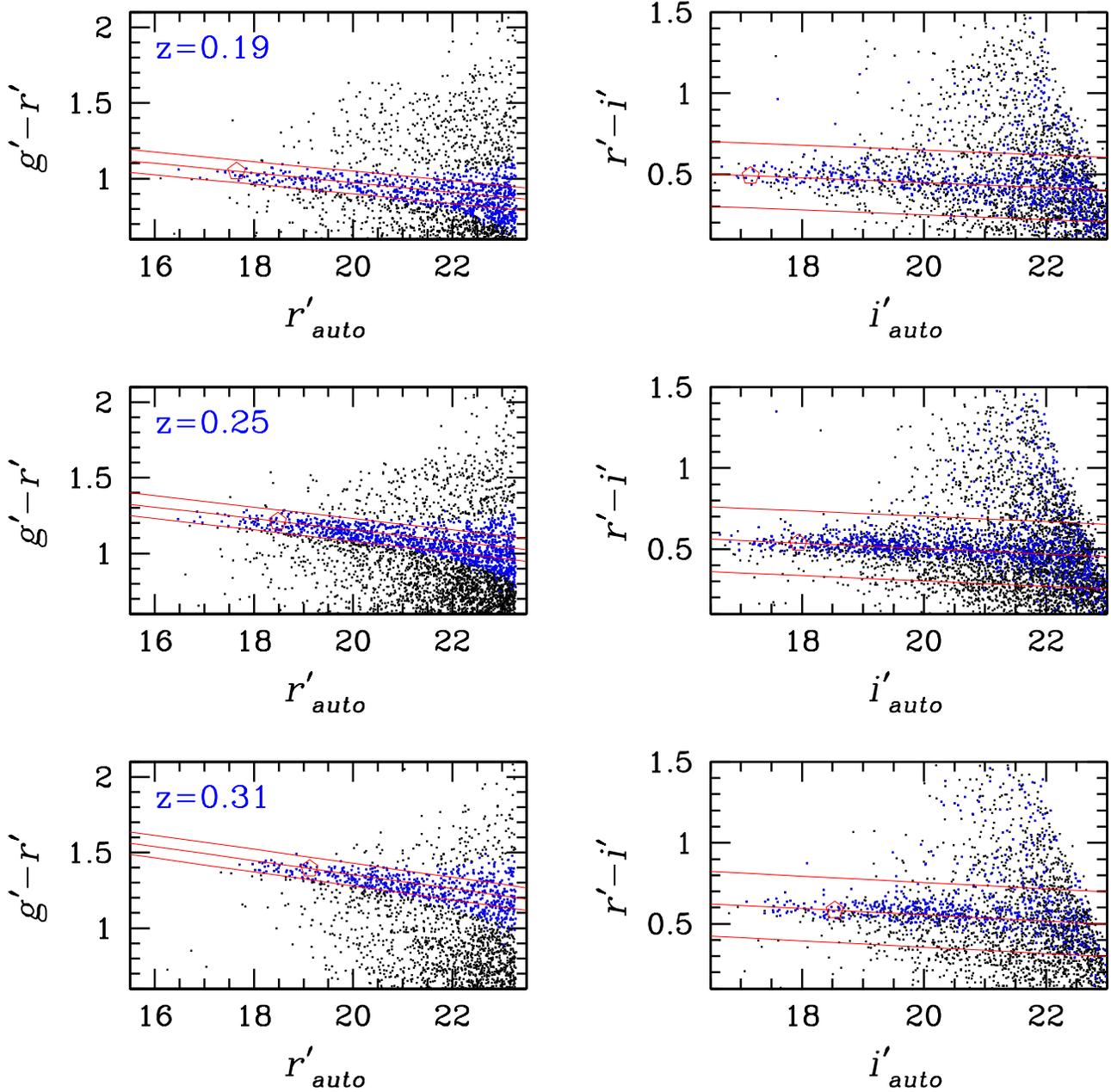}
 \caption{Stacked CMDs of galaxies within 0.5 Mpc around each cluster in $(g'-r')$ vs. $r'$ (black points in the left column) and $(r'-i')$ vs. $i'$ (black points in the right column) planes at three different redshifts. Pentagons indicate the position of $m^*$. Central solid lines in the $(g'-r')$ vs. $r'$ plots indicate the model CMR and the upper and lower bounds indicate the width of the colour slice, $\pm0.075$ mag. On the $(r'-i')$ vs. $i'$ plot, the central lines are the resulting CMR  from the same model that produces the $(g'-r')$ vs. $r'$ relation. The half width of the colour slice in $(r'-i')$ vs. $i'$ is 0.2 mag as indicated by the upper and lower solid lines (see text for explanation). Blue points are galaxies that belong to the red-sequence in $(g'-r')$ vs. $r'$. As we can tell, many of them do not fall on the red-sequence in the $(r'-i')$ vs. $i'$ plane, which shows the advantage of using two-colour information to select red galaxies. \label{ristacked}}
\end{figure*}

\subsection{Refinements to the Cluster Catalogue}
\subsubsection{High-z Contamination}

For any cluster candidate we detected, higher-redshift blue galaxies projected along the line-of-sight may have the same $(g'-r')$ colour as the red-sequence members of that cluster. This is demonstrated in Figure \ref{grritrack}, which shows the evolutionary tracks of $(g'-r')$ vs. $(r'-i')$ with redshift for three different populations produced by the model of \cite{galexev}. Squares are a population that formed 10.3 Gyr ago and have constant star formation. Pentagons represents a 8 Gyr old single-burst model and triangles a 10.3 Gyr single-burst model. The lines are to guide the eyes. For the two old models, galaxies in the redshift range $0.4<z<0.9$ all have similar $(g'-r')$ colour. However, they have different $(r'-i')$ colours. Therefore, $(r'-i')$ colour can effectively eliminate high redshift galaxies from the red-sequence galaxies at a lower redshift. This is further demonstrated in Figure \ref{ristacked}, which shows the stacked CMDs of galaxies within 0.5 Mpc around each cluster in $(g'-r')$ vs. $r'$ (black points in the left column) and $(r'-i')$ vs. $i'$ (black points in the right column) planes at three redshifts.  Pentagons indicate the position of $m^*$. Central solid lines in the $(g'-r')$ vs. $r'$ plots indicate the model CMR and the upper and lower bounds indicate the width of the colour slice, $\pm 0.075$ mag. On the $(r'-i')$ vs. $i'$ plot, the central lines are the resulting CMR from the same model that produces the $(g'-r')$ vs. $r'$ relation. The half width of the colour slice in $(r'-i')$ vs. $i'$ is  0.2 mag (justified below) as indicated by the upper and lower solid lines. Blue points are galaxies that belong to the colour slice in the $(g'-r')$ vs. $r'$ plane. In the $(r'-i')$ vs. $i'$ plane, most of those galaxies still fall on the red-sequence at low redshift, but as we go to higher redshift, many of them fall off the red-sequence even at bright magnitudes. Therefore, only galaxies that are on the red-sequence in both  $(g'-r')$ vs. $r'$ and $(r'-i')$ vs. $i'$ planes are considered as potential red-sequence cluster
members. Note we only use $(r'-i')$ colour to eliminate obvious background contamination; therefore we take a wide colour slice in $(r'-i')$ vs. $i'$ so that we do not lose red-sequence galaxies due to the uncertainty on the $(r'-i')$ colour and any slight mismatches between the model $(g'-r')$ and $(r'-i')$ colour slices. This technique has been used in \cite{andreon}.

\subsubsection{Centres}
As described above, the centre of each cluster is determined by the position at which the number of red-sequence galaxies within a radius of 0.5 Mpc selected using $(g'-r')$ vs. $r'$  is maximised. We refine the determination of the centre in several ways. We divide the region within a radius of 0.5 Mpc  around each cluster further into smaller grids of size $\sim0.06$ Mpc and count galaxies that belong to the red-sequence in both $(g'-r')$ vs. $r'$ and $(r'-i')$ vs. $i'$ planes in a circle of radius $\sim0.1$ Mpc. We also calculate the luminosity-weighted centre using galaxies that belong to the red-sequence in both $(g'-r')$ vs. $r'$ and $(r'-i')$ vs. $i'$ planes in a circle of radius $\sim0.5$ Mpc around each cluster. We compare the centres obtained in both ways with the position of the brightest cluster galaxy  in each cluster. For some rich, symmetric systems, the three centres agree with each other well. In the calculations later in the paper, we use the luminosity-weighted centre.

\subsubsection{Redshift}
As described in \S \ref{mod} (Figure \ref{CMgr}), the redshift interval between adjacent colour slices is 0.03. To refine the redshift estimate of each cluster candidate, we insert two more colour slices between two existing adjacent ones. To determine, for each cluster, which redshift the model CMD fits the observed red-sequence the best, two criteria are used: 1) we count the number of red-sequence galaxies within a circle of 0.5 Mpc in radius around the luminosity-weighted centre of that cluster; 2) we calculate the deviation of the colours from the model CMR for all red-sequence galaxies that are brighter than  $m^*+2$. We do this for each cluster in several adjacent colour slices and determine which one gives the highest number count and least deviation in colour. If these two criteria give the same optimal colour slice, then the redshift of that colour slice is assigned to the cluster. If these two select two different optimal colour slices, the one that gives the least deviation is chosen only if at least five galaxies are used in the fit and the deviation is significantly smaller than that from the slice that gives the highest count. We check how well this works by stacking all clusters at the same redshift and plotting the observed $(g'-r')$ vs. $r'$ relation against the model. 

 Finally, we interpolate the CMD in the $(r'-i')$ vs. $i'$ plane for the corresponding new redshift. A new centre is calculated for each cluster using the galaxies that fall on the red-sequence defined by the finely-interpolated colour slices in both the $(g'-r')$ vs. $r'$ and $(r'-i')$ vs. $i'$ plane. The significance of each cluster is re-estimated around the new centre.

\section{Cluster Properties}\label{prop}
\subsection{Richness}\label{pprich}
\begin{figure}
\leavevmode \epsfysize=8cm 
\leavevmode \epsfysize=8cm \epsfbox{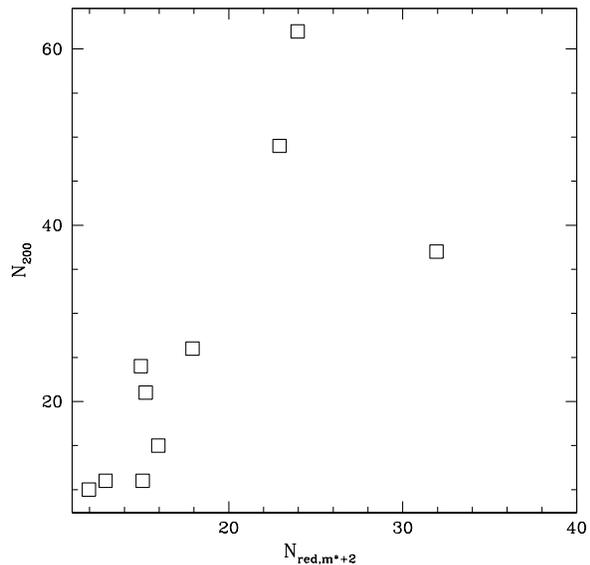} 
\caption{Correlation between our $N_{red,m^*+2}$ measured within 0.5 Mpc and the number of cluster members within $r_{200}$ from the MaxBCG sample \protect \citep{MaxBCG}.  
\label{p5N200}}
\end{figure}

Since we used the number of red-sequence galaxies brighter than $m^*+2$, and within only 0.5 Mpc from the cluster centre, in the cluster detection to reduce the noise, throughout this paper we use this number, denoted as $N_{red,m^*+2}$, to indicate the richness of the clusters in our sample. However, the typical extent of a cluster, $r_{200}$ (the radius within which the mean density is 200 times the critical density), is larger than 0.5 Mpc; therefore, to get some idea of how our  $N_{red,m^*+2}$ corresponds to the more commonly used richness indicator, $N_{200}$ (the number of cluster members within $r_{200}$), we plot $N_{red,m^*+2}$ vs. $N_{200}$ for the ten clusters that are both in our sample and the  MaxBCG catalogue \citep{MaxBCG} at this redshift range in Figure \ref{p5N200}.  Based on the relation $r_{200}=0.26N^{0.42}_{200}$Mpc \citep{johnston,hansen07}, the $r_{200}$ for clusters that are in the lower left corner of Figure \ref{p5N200} is about 0.8 Mpc; thus, their $N_{red,m^*+2}$ as measured within 0.5 Mpc is only slightly lower than  $N_{200}$. As one goes to richer clusters, the difference between $N_{red,m^*+2}$ and $N_{200}$ becomes larger. Nonetheless, we can approximately scale $N_{red,m^*+2}$ to  $N_{200}$ using Figure \ref{p5N200}.

\begin{figure}
\leavevmode \epsfysize=8cm \epsfbox{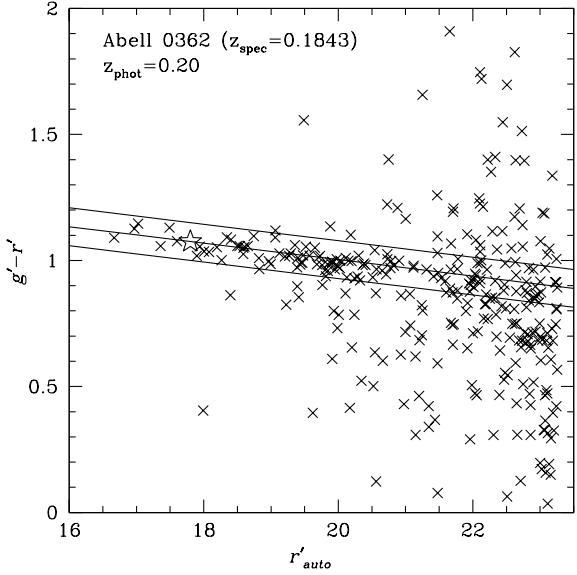}
\caption{The CMD of the richest cluster in our catalogue (Abell 0362 \protect \citealt{abell0362}) before background subtraction. Solid lines indicate the colour slice it belongs to with the star indicating the position of $m^*$,  and crosses are galaxies within a radius of 0.5 Mpc from the centre.\label{mostmassive}}
\end{figure}

The richest cluster in our catalogue has a $N_{red,m^*+2}$ of 46, a known Abell  cluster (Abell0362) with Richness Class 1 and $z=0.1843$ \citep{abell0362}. Its CMD is shown in Figure \ref{mostmassive}. Solid lines indicate the colour slice it belongs to with the	 star indicating the position of $m^*$, and crosses are galaxies within a radius of 0.5 Mpc from the centre (before background subtraction).

\subsection{Mass Estimates}\label{ppmass}
\begin{figure}
\leavevmode \epsfysize=8cm \epsfbox{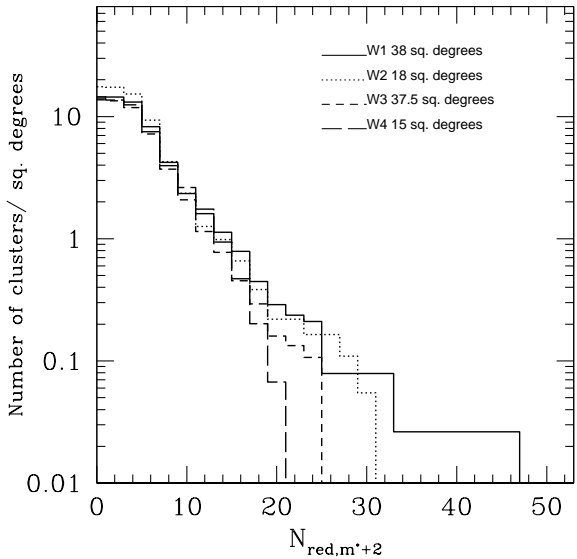}
\caption{Cumulative surface density of the clusters we detected as a function of $N_{red,m^*+2}$  in the redshift range $0.17\leq z \leq 0.36$, in all four wide fields. Each line style represents one of the four wide fields.\label{richsf}}
\end{figure}

To get some idea of how massive the clusters in our sample are, we compare the surface density of the clusters in our catalogue with that from the  Hubble Volume light cone. Based on the MS (spheres) cluster catalogue \citep{hubble}, there are 1.26 clusters  per square degree with mass greater than $1.4\times 10^{14}M_{\odot}$ in the redshift range of $0.17\leq z \leq 0.36$, and this surface density corresponds to that of the  clusters with $N_{red,m^*+2}\geq 12$ in our catalogue (as can be seen in Figure \ref{richsf}).

We have calculated the projected correlation function, $\omega(\theta)$, of the clusters in each of
the four wide fields separately, following the method of
\citet{ls93}.  We select all clusters in our sample with $0.17\leq z \leq 0.36$,
and a richness $N_{red,m^*+2}>10$, which leaves us with
74, 31, 57 and 32 clusters in the fields W1 through W4, respectively.
This richness limit is slightly poorer than that
adopted for most of the analysis in this paper in order to have
sufficient statistics for the following analysis.  This means our
results in the paper actually correspond to systems slightly {\it more}
massive than the limits given here.

\begin{figure}
\leavevmode \epsfysize=8cm 
\leavevmode \epsfysize=8cm \epsfbox{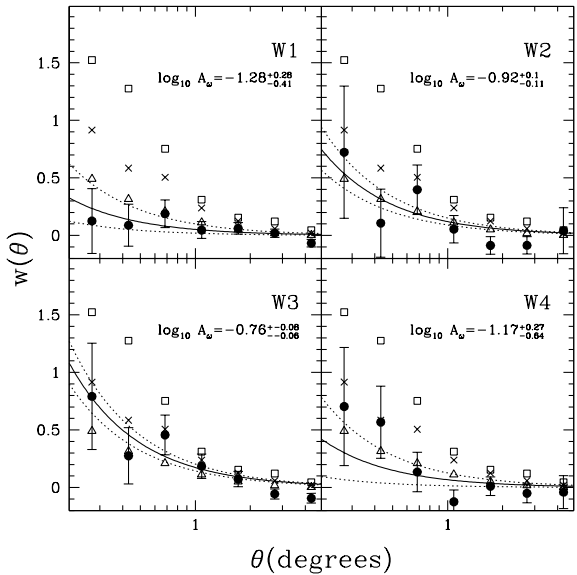} 
\caption{For each of the four CFHTLS survey fields, we show the angular
  correlation function for clusters with 
  $N_{red,m^*+2}>10$, as the solid points with Poisson error bars.  The
  solid line is the best fit power law
  $\omega(\theta)=A_\omega\theta^{1-\gamma}$, with $\gamma=2.15$
  fixed.  The dotted lines show the 95 per cent confidence limits.  The
  triangles, crosses and open squares are theoretical angular correlation functions
  for clusters with $M=8.0\times10^{13} M_\odot$,
  $M=3\times10^{14} M_\odot$, and $M=5\times10^{14} M_\odot$, respectively.  These were
  computed from the Hubble volume simulation over a similar redshift
  range as the data.  
\label{cf}}
\end{figure}

The correlation function is shown in Figure \ref{cf}. The random point distribution we use
for comparison does not account for the intrinsic size of our detection
filter; this means we will not measure the correlation function
accurately on angular sizes of about $\theta\leq 0.1$ degree.  We therefore
fit the data with a power law function
$\omega(\theta)=A_\omega\theta^{1-\gamma}$ over the range $0.2^o<\theta <2^o$.  We do not have enough data
to measure both the amplitude and slope, so we fix $\gamma=2.15$ as
measured locally for clusters \citep{G+02}.  We find the best fit
amplitude $A_\omega$ by maximizing the likelihood function as given in
\citet{G+02}, and take the 95 per cent confidence limits to be the
value where the relative likelihood is 0.1.  These values are given in
each panel in Figure \ref{cf}.  There is considerable variation from
field to field, to be expected since the areas are still fairly small.
The two fields with the most data, W1 and W3, actually represent the
least- and most-clustered data, respectively, though we note that only the W3
data actually puts useful constraints on the clustering amplitude.

Deprojecting the angular correlation function using the Limber equation
and assuming our standard cosmology, and including a Gaussian fit
to the observed $N(z)$ distribution (see Figure~\ref{zdistri}), we find
that the amplitude corresponds to a physical correlation length of
$r_\circ\sim 41$ Mpc for W3, and $23.8$ Mpc in the least-clustered field,
W1.  This range of correlation lengths in turn corresponds to a 
space density of $n_c=(0.2-4.6)\times10^{-6}$ (Mpc)$^{-3}$ \citep[e.g.][]{colberg}. We use the Millennium simulation \citep{ms} output at
$z\sim 0.3$, and find that this space density corresponds to 
clusters with mass $M\geq 1.0\times 10^{14}M_{\odot}$ (W1) and $M\geq 4.1\times 10^{14}M_{\odot}$ (W3).

As an alternative to deprojecting the angular correlation function, we
also compare our correlation function directly with that of the Hubble volume simulation
\citep{hubble}, by selecting clusters in the full-sky light cone,
within a similar redshift range as the data and in discrete bins of total mass.
We then compute the correlation function in the same way as for the
data; the results are shown in Figure \ref{cf} as triangles
($M>8.0\times10^{13} M_\odot$), crosses
($M>3.0\times10^{14} M_\odot$) and squares ($M>5.0\times10^{14}
M_\odot$).  In all fields except W1, the comparison with the data
suggests our sample with $N_{red,m^*+2}\geq 10$ is limited at about $M\geq 2\times10^{14} M_\odot$, consistent with the deprojection analysis above.

The CFHTLS W1 field also overlaps with the XMM-LSS field and we recovered, in the
redshift range we are interested in, all 9 X-ray confirmed clusters in
the XMM-LSS archive \citep{pacaud}. These clusters are all
low-richness clusters in our sample. Their X-ray temperatures range
from 1 keV to 3 keV.   From the observed mass-temperature
relation of galaxy clusters \citep[e.g.][]{evrard96,T_mass}, the mass
of a cluster with a X-ray temperature of $\sim 2$ keV is $\sim 1\times
10^{14}M_{\odot}$. This again indicates that the clusters in our sample
have masses of $M>1\times10^{14} M_\odot$. 

Throughout this paper, unless otherwise specified, the analysis is
carried out using the subset of 127 clusters that satisfy
$N_{red,m^*+2}\geq 12$. We emphasize that such a richness-selected
sample does not correspond exactly to a mass-selected one, and thus a
mass limit cannot be precisely defined.   Nonetheless the analyses above
show that our results are applicable to fairly massive clusters, with
$M>1\times 10^{14}M_{\odot}$; the 
sample is not dominated by low-mass groups.

\subsection{Redshift Accuracy}\label{ppz}
The redshifts we initially assign to our clusters are based on the photometric data. In this section, we assess how good our photometric redshift is.

\begin{figure}
\leavevmode \epsfysize=8cm \epsfbox{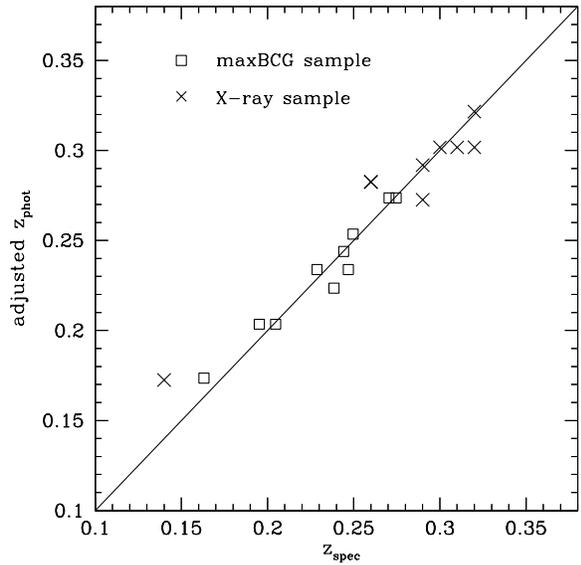}
 \caption{Comparison of our estimated photometric redshift with the spectroscopic redshift from the MaxBCG \protect \citep{MaxBCG} and X-ray sample \protect \citep{pacaud}. Squares are our $z_{phot}$ vs. $z_{spec}$ from maxBCG sample, with $z_{phot}$ corrected by a constant shift of 0.0435. Crosses are our corrected $z_{phot}$ vs. the $z_{spec}$  of the X-ray confirmed clusters from XMM-LSS in W1 field. The two agree with each other well, providing an independent verification of our correction.
\label{xrayz}}
\end{figure}

As mentioned in \S \ref{pprich}, there are ten common clusters in our catalogue and the MaxBCG catalogue \citep{MaxBCG}. For those clusters, we compare the spectroscopic redshifts of those BCGs with the photometric redshifts we assigned to those clusters. This comparison indicates that our photometric redshift is systematically lower, by a constant amount of 0.0435. Therefore, we apply this constant shift to the  estimated photometric redshifts from our initial model. (Recalibration in this way is a standard part of the CRS method, \citealt{GY00}.) All redshifts quoted in this paper have been adjusted in this way. Squares in  Figure \ref{xrayz} show the comparison between the adjusted $z_{phot}$ and the $z_{spec}$ of MaxBCG clusters.

We also compare the adjusted photometric redshifts of the X-ray confirmed clusters (\S \ref{ppmass}) in our catalogue with their spectroscopic redshifts, shown as crosses in Figure \ref{xrayz}. The two agree with each other well, providing an independent verification of our correction. The rms of the difference between the $z_{photz}$ and $z_{spec}$ (including both the MaxBCG and X-ray samples) is $\sim$0.014, with very little bias.

\begin{figure}
\leavevmode \epsfysize=8cm \epsfbox{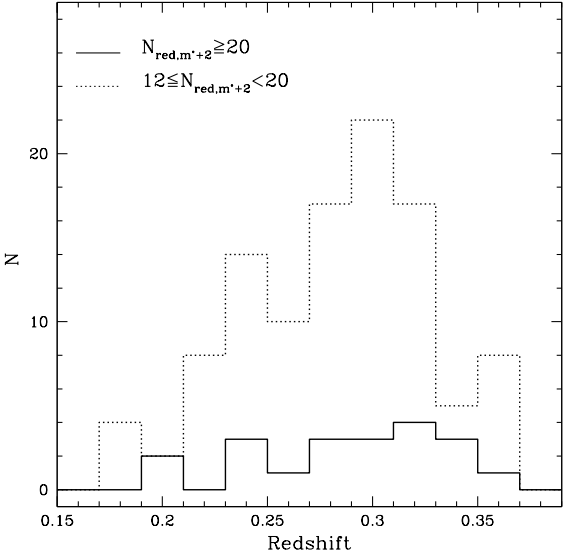}
\caption{Redshift distribution of clusters in our sample, split into two subsamples based on $N_{red,m^*+2}$. The solid histogram represents the distribution of the subsample of clusters with $N_{red,m^*+2}\geq 20$, and the dotted histogram represents those with $12\leq N_{red,m^*+2}< 20$. The distribution is smooth with redshift, peaking at around $z\sim 0.3$.\label{zdistri}}
\end{figure}

Figure \ref{zdistri} shows the distribution of clusters as a function of the adjusted $z_{phot}$ in our sample, split into two subsamples based on $N_{red,m^*+2}$. The solid histogram represents the distribution of the subsample of clusters with $N_{red,m^*+2}\geq 20$, and the dotted histogram represents those with $12\leq N_{red,m^*+2}< 20$. The distribution is smooth with redshift, peaking at around $z\sim 0.3$.

\section{Local Comparison Sample}\label{local}
It is useful to extend our redshift baseline by examining a sample of local clusters, using our methods consistently. Therefore, we make use of the low-redshift galaxy group catalogue by Yang et al. (2007, hereafter Yang07), selected using a Friends-of-Friends algorithm from  the SDSS spectroscopic data \citep{sloan}.  We calculate  $N_{red,m^*+2}$ for these groups/clusters the same way as for our own cluster sample, except that the red-sequence galaxies are selected in $(u-r)$ colour to bracket the 4000\AA \ break, because of the lower redshift of the sample. We select a subset of clusters that have  $N_{red,m^*+2} \geq 12$, the same richness cut of our cluster sample. In order to carry out the calculations (e.g. background subtraction) exactly the same way as for our own sample, we only choose clusters that are in a contiguous region of the survey, and this limits the number of clusters in our local comparison sample to 22, at $0.08<z<0.09$.

In addition, we repeat the same measurement for the rich $z=0.023$ cluster, Coma, using the same SDSS data, with the red-sequence galaxies selected using $(u-g)$ colour to bracket the 4000\AA \ break at its redshift. Note Coma has a  $N_{red,m^*+2} \sim 30$ and $N_{200}\sim 100$ (see \S \ref{pprich}), and therefore is   $\sim 4$ times richer than the typical clusters in our sample and the Yang07 sample.

\section{Luminosity Function Construction}\label{rel}

In this paper, we focus on the $r'-$band red-sequence luminosity function (LF) in the core regions of clusters over the redshift range of $0.17\leq z \leq 0.36$. We choose the  $r'$ band because it has the deepest photometry, and it is a red band which is less sensitive to recent star formation than bluer bands. We divide our cluster sample into three redshift bins  and stack all clusters in each bin to obtain a composite red-sequence LF, to reduce the noise due to the uncertainty in cluster membership determination and cluster-to-cluster variation. The width of the redshift bin is chosen in a way that the number of clusters in each bin is roughly the same, to give similar statistics.

$k$-corrections or $(k+e)$ corrections are applied when necessary. The corrections are calculated using the same old, single-burst \cite{galexev} model used to define the CMR, and are magnitude dependent. The $k$-correction is about $0.2-0.5$ mag at $z<0.27$ (the two lower redshift bins), and about $0.5-0.7$ mag at $z\sim 0.36$ (the highest redshift bin). The $(k+e)$ correction is less than $\sim 0.2$ mag at $z<0.36$.

\begin{figure*}
\leavevmode \epsfysize=18cm
\epsfbox{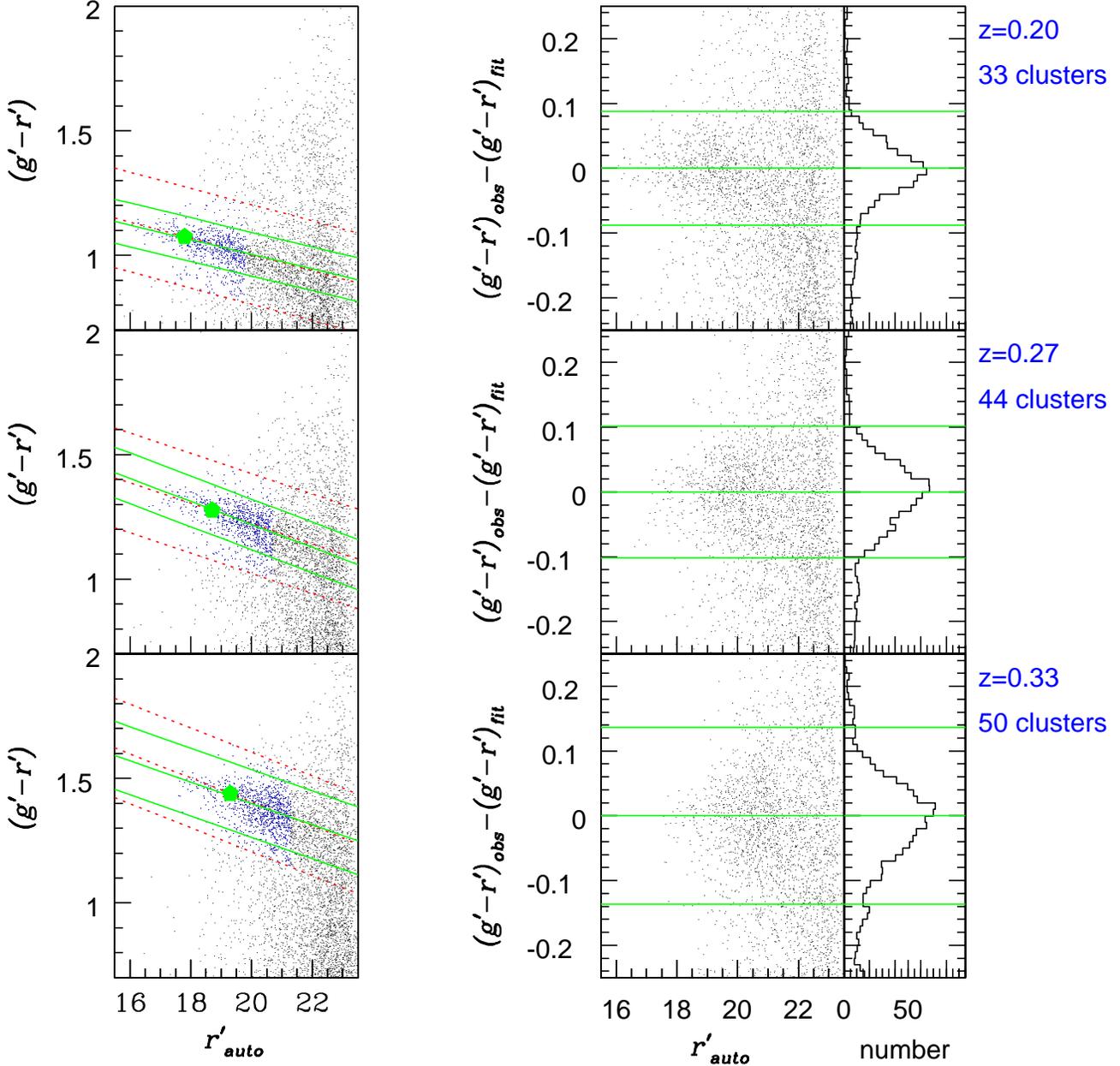}
\caption{Left panel: Composite CMDs of galaxies that are within a radius of 0.5 Mpc from the cluster centre at three different redshifts. The dashed lines indicate $\pm 0.2$ mag from the modelled $(g'-r')$ vs. $r'$ relation. Central solid  lines are the new best-fit CMR, based on galaxies that are brighter than $m^*+2$  ($m^*$ is indicated by the pentagon), shown as the blue points. The upper and lower solid lines are the 2$\sigma$ bounds. Right panel: in the left column is the CMD with the best-fit CMR subtracted, and in the right column the histograms show the distribution of galaxies relative to the fitted CMR down to $m^*+2$. The upper and lower solid lines in each panel are the 2$\sigma$ bounds. See text for details.\label{CMDzbin} }
\end{figure*}

We construct a composite CMD in each redshift bin to define the red-sequence. For clusters in each colour slice we calculate the position of every galaxy in the CMD relative to $m^*$ in that colour slice, and shift those that fall onto the red-sequence defined by the wide $(r'-i')$ colour slice (to eliminate obvious foreground and background contamination) to the central redshift in both colour and magnitude. The first column in Figure \ref{CMDzbin} shows the composite CMDs of galaxies that are within a radius of 0.5 Mpc from cluster centres at three different redshifts. The dashed lines indicate $\pm 0.2$ mag from the modelled $(g'-r')$ vs. $r'$ relation. To more accurately define the red-sequence, we fit a new $(g'-r')$ vs. $r'$ relation using galaxies brighter than $m^*+2$ ($m^*$ is indicated by the pentagon) that are within $\pm 0.2$ mag from the model $(g'-r')$ colour (as indicated by the blue points). We divide those galaxies into magnitude bins of 0.5 mag, calculate the median of the colour distribution in each magnitude bin, and fit the medians as a function of magnitude to a straight line, indicated by the central solid lines. To examine the fit more closely, we subtract the fitted $(g'-r')$ vs. $r'$ relation and plot the relative position of each galaxy to the fitted CMR, which is shown in the second column in Figure \ref{CMDzbin}, with the histograms in the third column showing the distribution of the residuals (down to $m^*+2$). The resulting CMD is centred at zero colour-difference relative to the fit by construction. To calculate the width, $\sigma$, of the colour distribution around the fitted $(g'-r')$ vs. $r'$ relation, we mirror galaxies that are redder than the fitted CMR to avoid contamination from the blue cloud, and apply $3\sigma$ clipping. This $\sigma$ is calculated from galaxies that are brighter than $m^*+2$, and is not magnitude dependent. The upper and lower solid lines in each panel are the 2$\sigma$ bounds of the colour distribution of galaxies brighter than $m^*+2$.

Isolating red-sequence galaxies is non-trivial; both intrinsic scatter and photometric uncertainties can make it difficult to cleanly separate the red-sequence and blue galaxies. This is particularly a problem at faint magnitudes, where photometric errors are large, and the blue cloud dominates the population. Here we select red-sequence galaxies in several different ways and show how they affect the results. The four methods we use are :\\
1) Red\_$4\sigma$, where red-sequence galaxies are defined as those that are redder, but not more than $4\sigma$, than the best-fit CMR. The total is twice this number.\\
2) Red\_all, a slight variation of method Red\_$4\sigma$, where red-sequence galaxies are defined as all those that are on the red side of the best-fit CMR. Again, the total is twice this number. Method Red\_$4\sigma$ and Red\_all are motivated by \cite{barkhouse} and \cite{gilbanklf} (but in their work the red-sequence galaxies are not mirrored at the bright end).\\
3) P10\_$2\sigma$, where red-sequence galaxies are defined as those that have $>10$ per cent probability belonging to the colour slice defined by the best-fit CMR with a width of $\pm 2\sigma$ as indicated by the solid lines in Figure \ref{CMDzbin}. This method is consistent with how the subsample in each colour slice is selected (\S \ref{sub}) prescribed by \citealt{GY00}. Note each galaxy that satisfies  this criterion is counted as one galaxy, not weighted by this probability.\\
4) NP\_$2\sigma$, where red-sequence galaxies are defined as those that are completely contained within $\pm 2\sigma$ from the best-fit CMR.\\

To correct for the background contamination, the same strategy is applied to a sample of galaxies at 500 random positions within the four CFHTLS wide fields, for each finely-interpolated colour slice. They are shifted to each central redshift bin in the same way as for the cluster+field sample, and the average number count is used when subtracting the field contribution. The error bars are estimated assuming that the noise follows Poisson statistics, including the background subtraction. Note that the Poisson error on the background is negligibly small, because of the large number of random fields. We do not include the variance of the background field distribution in the error estimate, because here we are primarily interested in the average properties  of a large ensemble of clusters, not the cluster-to-cluster variation.

\section{Results}\label{results}

\begin{figure*}
\epsfbox{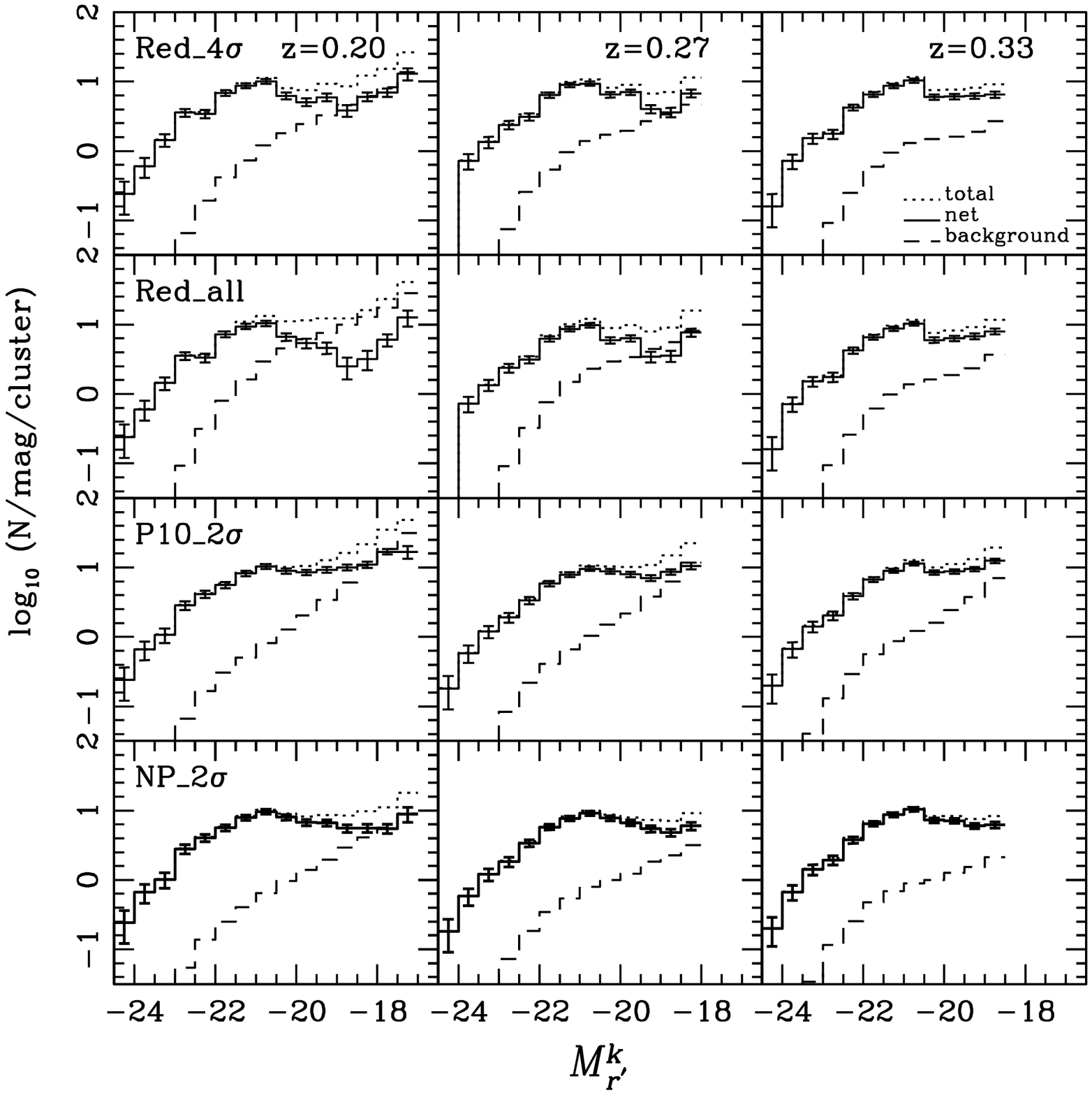}
\caption{Composite  $r'-$band red-sequence LFs of clusters with $N_{red,m^*+2}\geq 12$ in the projected radius range of $0\leq r <0.5$ Mpc, calculated using the four methods described in the text, shown in four rows respectively. The top row, Red\_$4\sigma$,  is our preferred method, as it minimizes the contamination from the blue cloud, and background galaxies. Magnitudes are $k$-corrected to rest frame. The dotted histograms are the cluster+field counts, the dashed histograms are background counts and the solid histograms with error bars are the background-subtracted LFs. See text for discussions on each method.\label{LFz3bin}}
\end{figure*}

\subsection{Red-sequence Luminosity Functions}\label{rslf}
\subsubsection{Our sample}\label{ourlf}

Figure \ref{LFz3bin} shows the  $r'-$band red-sequence LFs of all clusters with $N_{red,m^*+2}\geq 12$ in the range of projected radius $0\leq r <0.5$ Mpc, $k$-corrected to rest frame, calculated using the four methods described in \S \ref{rel}.  The depth of the CFHTLS data enables us to reach $M_{r'}\sim -17$ at $z\sim 0.2$, a limit that has never been probed before at this redshift for such a large sample of clusters.

In all panels in Figure \ref{LFz3bin}, the dotted histograms are the cluster+field counts, the dashed histograms are background counts, and the solid histograms with error bars are the background-subtracted LFs. The number count in each magnitude bin is normalized by the number of clusters contributing to that bin. The number of clusters in each redshift bin is 33, 44 and 50 from low to high redshift, and all clusters contribute fully to all but the faintest magnitude bin due to the completeness limit.

 At bright magnitudes, all four methods give consistent results, so it does not matter whether one uses only galaxies on the redder side of the CMR or not. However it makes a difference at faint magnitudes. In Figure \ref{LFz3bin}, comparing the net counts from the four methods at magnitudes fainter than $M^k_{r'}=-20.5$ (2 magnitudes fainter than $M^*$), P10\_2$\sigma$ gives the highest count. This is an overestimate because, at faint magnitudes, the error on the colour is larger; galaxies that are just outside the colour slice can still have a probability of $>10$ per cent belonging to the colour slice. Note that the LFs are background-subtracted; therefore this higher net count in method P10\_2$\sigma$ at faint magnitudes is not due to field contamination, but due to the contribution from the cluster blue cloud. Method NP\_2$\sigma$  suffers from the same problem as P10\_2$\sigma$ over the magnitude range $-20.5<M^k_{r'}<-18.5$, as can be seen from the net count shown in the bottom row. However, at the very faintest magnitudes, NP\_2$\sigma$ gives the lowest net count among the four methods. This is because, in this case, the width of the fixed colour slice is not wide enough compared to the error on the colour at this faint magnitude; thus it underestimates the number of red-sequence galaxies, due to the preferential scattering of the red-sequence galaxies out of the colour slice. This problem cannot be alleviated by using a wider colour slice because then it would significantly overestimate the number of red-sequence galaxies at slightly brighter magnitudes given the above reasoning about method P10\_2$\sigma$.

On the other hand, method Red\_$4\sigma$ and its slight variation, Red\_all, mirror only galaxies on the redder side of the CMR, and thus do not include contributions from the cluster blue cloud that occupy regions blueward of the CMD (especially at faint magnitudes). Moreover, since method Red\_all takes account of all galaxies redder than the best-fit CMR, it does not underestimate the number of red-sequence galaxies regardless of the photometric errors. In method Red\_$4\sigma$, a cut of $4\sigma$ redder than the best-fit CMR is applied to eliminate high redshift galaxies with colours much redder than the best-fit CMR. This cut corresponds to $\sim 0.17-0.27$ mag from our lowest redshift bin to the highest, which is comparable to or greater than the $1\sigma$ error on the colour even at the very faintest magnitude ($\sim 0.2$ mag at $r'\sim 23$, see Figure \ref{grerrlinfit}); therefore, this cut is broad enough to not significantly underestimate the number of red-sequence galaxies scattered off the CMR due to photometric errors. Comparing the results from these two methods (first and second row in Figure \ref{LFz3bin}), we see that the net counts are consistent (solid histograms), but the background (dashed histograms) in method Red\_$4\sigma$ is significantly reduced compared to that in method Red\_all; therefore, we conclude that Red\_$4\sigma$ is the best, and  in the rest of the paper, we will use this method for all the analysis. 

The most outstanding feature of our red-sequence LF at $z\sim 0.2$ is a significant and broad dip starting at $M_{r'} \sim -20.5$. The  number of red-sequence galaxies reaches its maximum at $M_{r'}\sim -20.5$, and then decreases to 40 per cent of the maximum value at $M_{r'}\sim -18.5$, and comes back up at magnitudes fainter than that. At this redshift, $M_{r'}\sim -18.5$ corresponds to $r'\sim 22$, where the error on the colour and total magnitude is still small (see Figure \ref{grerrlinfit}); thus this inflexion is robust. This feature is also present at $z\sim 0.27$. It is hard to discern at $z\sim0.33$ due to the magnitude limit of the data and the fact that the error on the colour is getting large at the very faintest magnitudes. The dip in the LF probably suggests that the LF is made up of a mixture of two populations of red-sequence galaxies, possibly giant/regular elliptical and dwarf ellipticals (dEs): as can be seen in figure 1 of \cite{binggeli}, the LF of E and S0 galaxies peak at bright magnitudes and drop off at the faint end where the LF of dEs starts to increase. The disappearance of dEs over the magnitude range seen in our LF could mean that those dwarf galaxies on this mass scale are either disrupted (by, for example, tidal forces) or they merge into more massive galaxies.

Although the existence of dips in the red-sequence/early-type LFs has been established in many studies \citep[e.g.][]{popesso,barkhouse,mercurio06,sh96}, the depth of the dip and the faint-end slope of the LFs are still not well constrained, varying significantly in different studies. We examine this issure more closely in \S \ref{cplit}.

\begin{figure*}
\epsfysize=18.5cm
\epsfbox{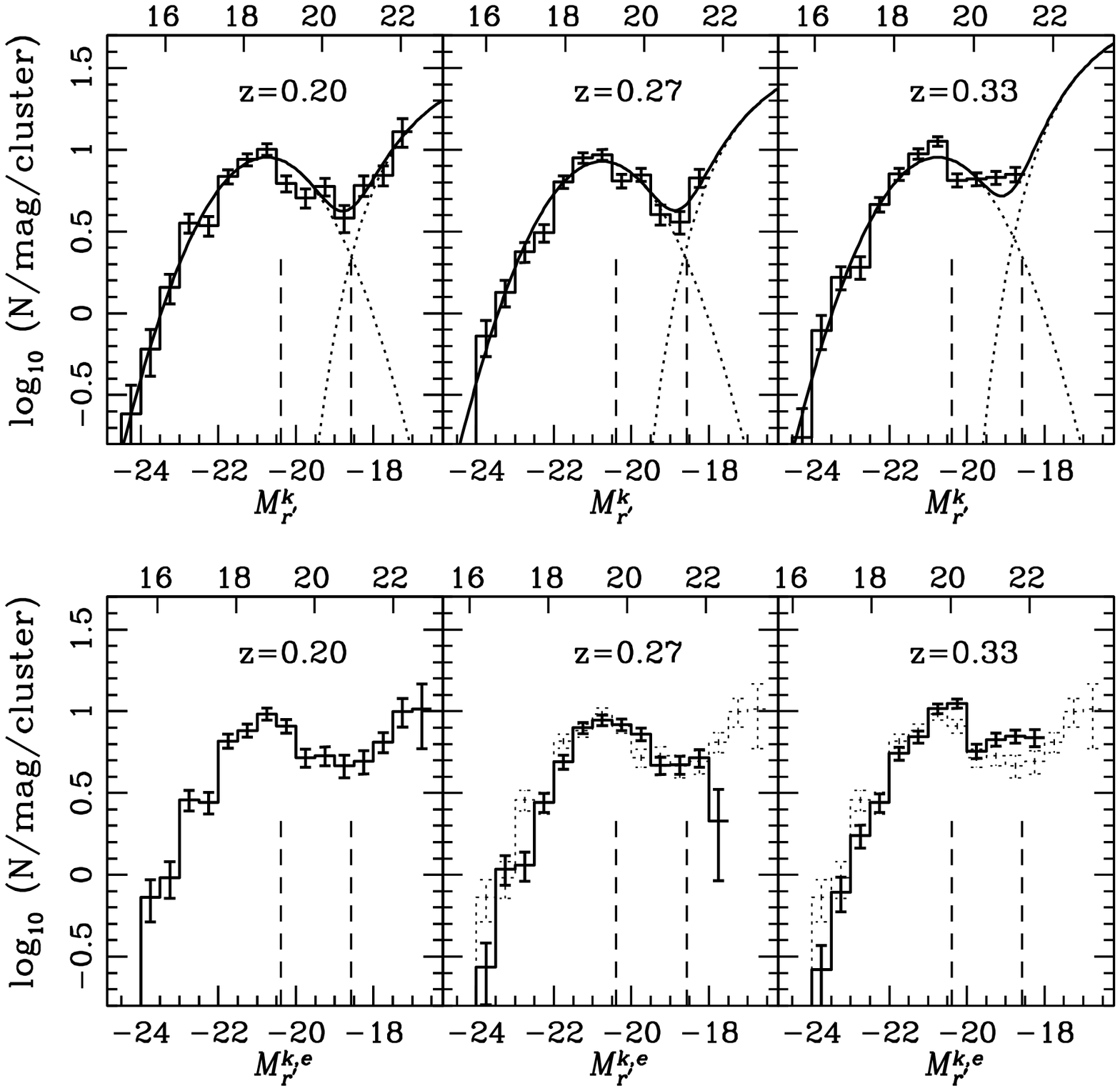}
\caption{Top panel: $k$-corrected, background-subtracted red-sequence LFs (histograms) obtained using method Red\_$4\sigma$ in three redshift bins with the best-fit Gaussian + Schechter functions overplotted as solid curves (dotted curves show the two components separately). For the lowest redshift bin, all 6 parameters in the fit are free; for the two higher redshift bins, the $M^*$ and $\alpha$ are fixed at the best-fit values from the lowest redshift bin. $k$-corrected rest-frame absolute magnitude and apparent magnitude are labelled on the bottom and top axes, respectively. The two vertical dashed lines indicate the division between dwarf and giant as in \protect \cite{lucia}. Bottom panel: LFs with $(k+e)$-correction applied. Solid histograms represent $(k+e)$-corrected LF at each redshift. The one in the lowest redshift bin is plotted for reference on top of that in the two higher redshift bins as dotted histograms. Top and bottom axes label the apparent magnitude and $(k+e)$-corrected rest-frame absolute magnitude, respectively. Again, the vertical dashed lines indicate the division between dwarf and giant. 
\label{schfit}}
\end{figure*}

 In most cases, instead of fitting a single Schechter function, double Schechter functions: one for the bright end and one for the faint end; or Gaussian (bright end) + Schechter (faint end) functions are used to try to accurately represent the shape of the LFs with dips. In our two lowest redshift bins, we find that double Schechter functions do not provide a good fit to the LF (without removing the brightest cluster galaxies).  Instead, a Gaussian + Schechter function fit describes the shapes of our LFs better. For the highest redshift bin, due to the limit of the data, it is hard to tell whether the inflexion exists; therefore a single Schechter function provides an acceptable fit as well.  In the lowest redshift bin, all six parameters in the Gaussian + Schechter function are free; but in the two high redshift bins, the data are not adequate to constrain the faint end and therefore we fix the $M^*$ and $\alpha$ at the best-fit values from the lowest redshift bin. Note we do not explore the degeneracy among the parameters as that is not our purpose here; we only seek a set of parameters that accurately reflect the shape of our LFs. The best-fit parameters 
 of our LFs obtained using method Red\_$4\sigma$ are listed in Table \ref{schparam} and the fits are plotted as solid curves over the $k$-corrected and background-subtracted LFs (histograms) in the top panel in Figure \ref{schfit}. Within the uncertainty, the fit to the LF in the lowest redshift bin can also fairly represent the LFs in the two higher redshift bins. Note the fits are for $k$-corrected LFs, but given the low redshift probed here, the evolutionary correction is small; thus this implies no strong evolution in this redshift range.

In the bottom panel of  Figure \ref{schfit}, we plot our $(k+e)$-corrected LFs. Solid histograms represent the LF in each redshift bin, and the one from the lowest redshift is overplotted as dotted histogram in the two high redshift bins. There does not seem to be a strong evolution above that of passive evolution.  In each panel, the two vertical dashed lines indicate the division between dwarf and giant as in \cite{lucia}. Galaxies brighter than the magnitude indicated by the line on the left are considered giants; and galaxies between the two lines are defined as dwarfs (the dwarf-to-giant ratio, DGR, will be discussed later on). 

\begin{table}
\caption{Best-fit parameters of the LFs of our cluster sample constructed using method Red\_$4\sigma$. Bracketed values are fixed from the lowest redshift bin.}
\begin{tabular*}{0.4812\textwidth}{|cc|cc|cc|cc|} \hline
 \hspace {18pt} \vline &  Gaussian  & &  Schechter & \\
\hline
 \ \  $z$ \hspace {5.3pt} \vline &    \ $<M_{r'}>$\ \ \ \  $\sigma$ \ \ \  \ $\phi_{gau}$ &&\hspace{1pt} $M^*_{r'}$\hspace{22pt} $\alpha$\hspace{23pt} $\phi$ \\
\hline
      0.20\ \ \vline  &\   -20.75 \ \  \ \ 1.3 \ \  \ \ 9.0&&\hspace{1pt} -17.75\hspace{13.2pt} -1.15\hspace{10.5pt}  22.5 \\
\hline
      0.27 \ \vline & \   -20.75 \ \ \ \ 1.3 \ \ \ \  8.5&& (-17.75)\hspace{7pt} (-1.15)\hspace{7pt} 26.5\\
\hline
    0.33 \ \vline  &\   -20.75 \ \ \ \  1.3  \ \ \ \ 9.0&& (-17.75)\hspace{7pt} (-1.15)\hspace{7pt} 50.0 \\
\hline

\end{tabular*}\label{schparam}
\end{table}

\subsubsection{Low-redshift comparison}\label{lfzev}
\begin{figure}
\leavevmode \epsfysize=8cm
\epsfbox{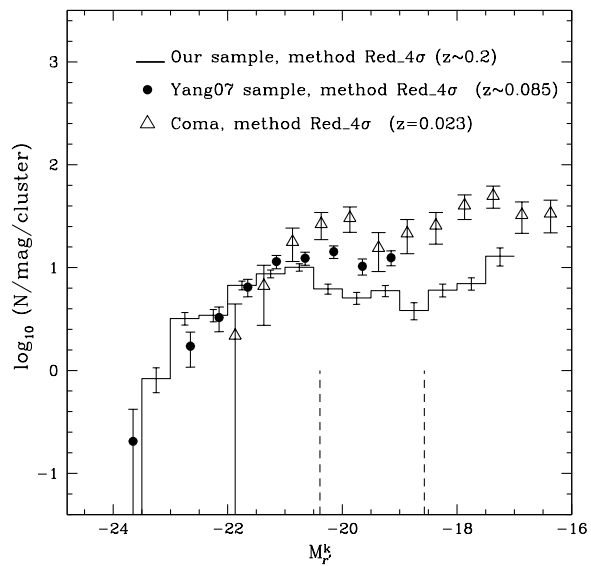}
\caption{$r'-$band red-sequence LFs at different redshifts, passively evolved to $z=0.2$. Magnitudes are $k$-corrected to rest-frame at $z=0$. The solid histogram is the red-sequence LF of our sample at $z\sim 0.2$, with the brightest cluster galaxies excluded. Solid points and triangles are the red-sequence LF of the Yang07 sample ($z\sim 0.085$, \S \ref{local}) and Coma ($z=0.023$), respectively. All LFs are calculated using our method Red\_$4\sigma$, and normalized to give the same cumulative number count down to $M^k_{r'}=-20.5$.  The two vertical dashed lines show the approximate division between dwarf and giant as in \protect \cite{lucia}. See text for details.\label{litfit_z}}
\end{figure}

As seen in the above section, there does not seem to be a strong evolution of the red-sequence LFs within our sample at $0.2\lsim z\lsim0.4$. In this section, we compare the LFs of our sample with that of the Yang07 sample and the intensely studied local rich cluster, Coma. In Figure \ref{litfit_z}, the solid histogram is the red-sequence LF of our sample at $z\sim 0.2$, with the brightest cluster galaxies removed (for easier comparison with the literature later). Solid points and triangles are the red-sequence LFs of the Yang07 sample ($z\sim 0.085$, \S \ref{local}) and Coma ($z=0.023$), respectively, calculated using our method Red\_$4\sigma$, and passively evolved to $z=0.2$. To compare the shapes of these LFs, in Figure \ref{litfit_z} they are all normalised to give the same total number count down to $M_{r'}= -20.5$.  Since these LFs are constructed using the same method, and the filter combinations used at these redshifts bracket the 4000\AA \ break in a similar way, the differences in their faint ends are not likely due to method differences, but genuine, suggesting a steepening of the faint end with decreasing redshift (but note that Coma is  $\sim 4$ times richer than the typical clusters in our and the Yang07 samples, and we will discuss this in \S \ref{richdgr}).

\subsection{Red-sequence Dwarf-to-Giant Ratio (DGR)}\label{secdgr}
The DGR is the ratio of the number of faint galaxies to bright ones, and thus is commonly used as a simple indicator of the shape of the
LF. Since red-sequence galaxies represent the population that have had their star formation shut off, the evolution of red-sequence DGR traces the history of the quenching of the star formation.

\subsubsection{DGR of Our Cluster Sample}\label{ourcl}
\begin{figure*}
\epsfbox{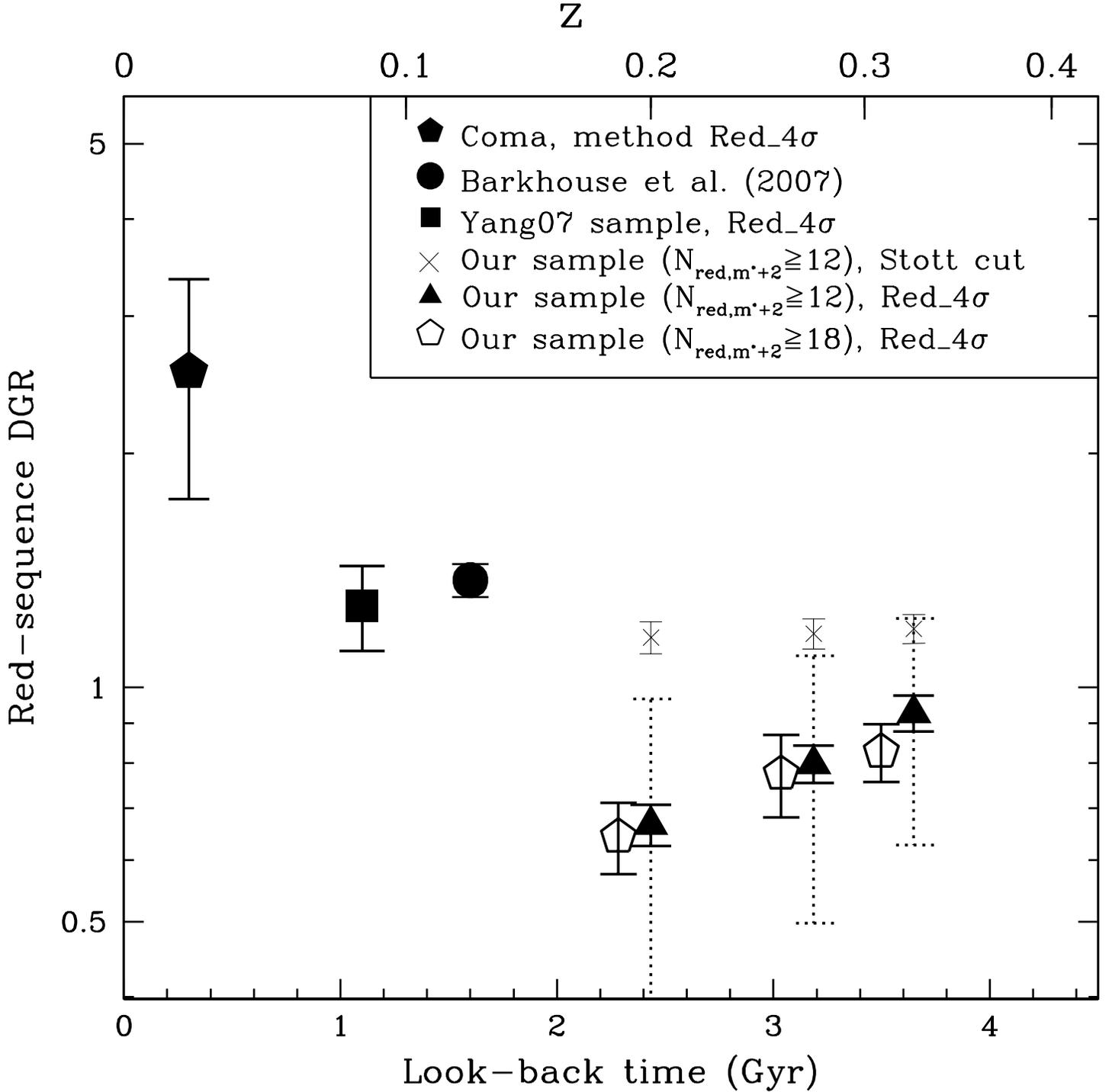}
 \caption{Red-sequence DGRs as a function of look-back time and redshift. Triangles are the DGRs of our cluster sample with $N_{red,m^*+2}\geq 12$, in the radius range of $0\leq r<0.5$ Mpc, calculated using method Red\_$4\sigma$. The solid error bars indicate uncertainties on the sample average DGRs,  while dotted error bars represent typical errors on the DGR of a single cluster whose DGR is similar to the ensemble average. Open pentagons are the DGRs of a rich subset of our cluster sample with $N_{red,m^*+2}\geq 18$ (discussed in \S \ref{richdgr}), plotted with slight offset in redshift for clarity. The solid square and pentagon are the red-sequence DGR of the Yang07 sample and Coma, calculated using method Red\_$4\sigma$. The filled circle is from \protect \cite{barkhouse}. Crosses are the DGR of our sample, calculated using the \protect \cite{stott} colour cut. See text for more discussion.\label{DGRp5Mpcbus}}
\end{figure*}

To make it easier to compare with previous works, here we $k+e$ correct the apparent magnitude to rest frame at $z=0$, and adopt the definition of
dwarf and giant used by \cite{lucia}, i.e. galaxies
brighter than $-20$ in rest-frame $V$ in the Vega system are considered as
giant and those between $-20$ and $-18.2$ are defined as dwarf. We
convert the definition of dwarf and giant in $V$  to our rest-frame $r'$ magnitude using the $(V-r')$ colour produced by the same old, single-burst model described in \S \ref{mod}.  
The triangles in Figure \ref{DGRp5Mpcbus} are the red-sequence DGRs of our cluster sample with $N_{red,m^*+2}\geq 12$, in the radius range  $0\leq r<0.5$ Mpc, calculated using method Red\_$4\sigma$.  The bottom and top axis labels show the look-back time and redshift respectively. The small error bars on our data points are estimated assuming Poisson statistics, without including the variance of the background field distribution, because what we are after is the average behaviour of the large sample. To get an idea of how large the error bar would be on the DGR of a single cluster, we calculate the uncertainty on the DGRs for a few individual clusters that have DGRs similar to the ensemble average,  taking into account the noise on the mean background and the variance of the distribution of the 500 random fields. The typical uncertainties are plotted as dotted error bars on the triangles in  Figure \ref{DGRp5Mpcbus}.  We also find a wide cluster-to-cluster variation, and 8 out of 127 of our clusters have $DGR>2$. We will explore this  further in the following discussion section.

\subsubsection{Evolution with Redshift}\label{zev}
To explore the redshift evolution of the  red-sequence DGR, we also calculate the DGR for the local comparison sample  and Coma (\S \ref{local}), shown as the square and solid pentagon in Figure \ref{DGRp5Mpcbus}. The DGRs we have calculated consistently  at different redshifts show no strong evolution in the redshift range $0.2\lsim z\lsim0.4$, but a rapid increase from $z\sim 0.2$ to $z\sim 0$. Assuming that the bright end of the LF has not evolved significantly, our results imply that the number of dwarf galaxies has increased by a factor of $\sim 3$ over the last 2.5 Gyr. This rapid evolution since $z\sim 0.2$ is not inconsistent with the predictions from either single burst simple stellar population (SSP) or  quenched models by \cite{redfrac} through the measurement of the age of faint red galaxies in Coma using line indices (see their figure 15), although our results agree more with their prediction for the outskirt regions instead of the core regions. Despite the rapid increase in the number of dwarf galaxies, they do not contribute significantly to the growth of the stellar mass on the red-sequence, {\it {per cluster}}. Using the stellar mass-to-light ratios from the same simple model we used to construct the CMD, an increase in the DGR from $\sim 0.8$ to $\sim 2.5$ corresponds to a $\sim 15$ per cent increase in the  stellar mass on the red-sequence, {\it {per cluster}}.

\section{Discussion}\label{disc}

\subsection{Comparison with Literature}\label{cplit}
Studies on cluster red-sequence LF in the literature have used quite different colour cuts to select red-sequence galaxies. In this section, we compare our red-sequence LFs and DGRs with that in the literature. Given the redshift evolution shown above, we split the comparison into two redshifts.  The LFs in the literature are transformed into our $r'$ filter and passively evolved  to $z=0.2$, using the same SSP \cite{galexev} model described in \S \ref{mod}. Magnitudes are converted to the cosmology used in this work, if different.

\begin{figure*}
\leavevmode \epsfysize=8cm
\includegraphics[width=90mm]{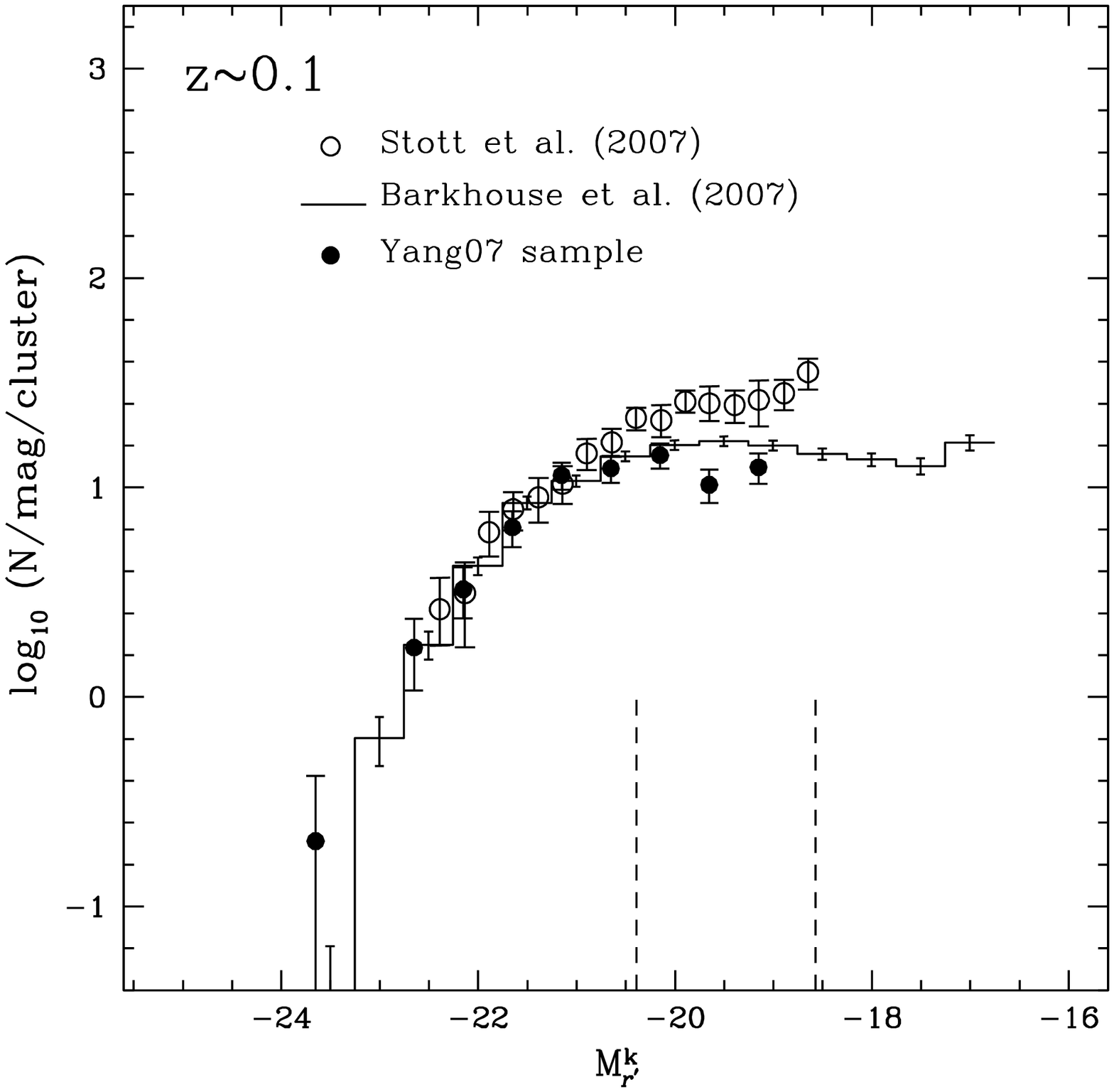}\includegraphics[width=90mm]{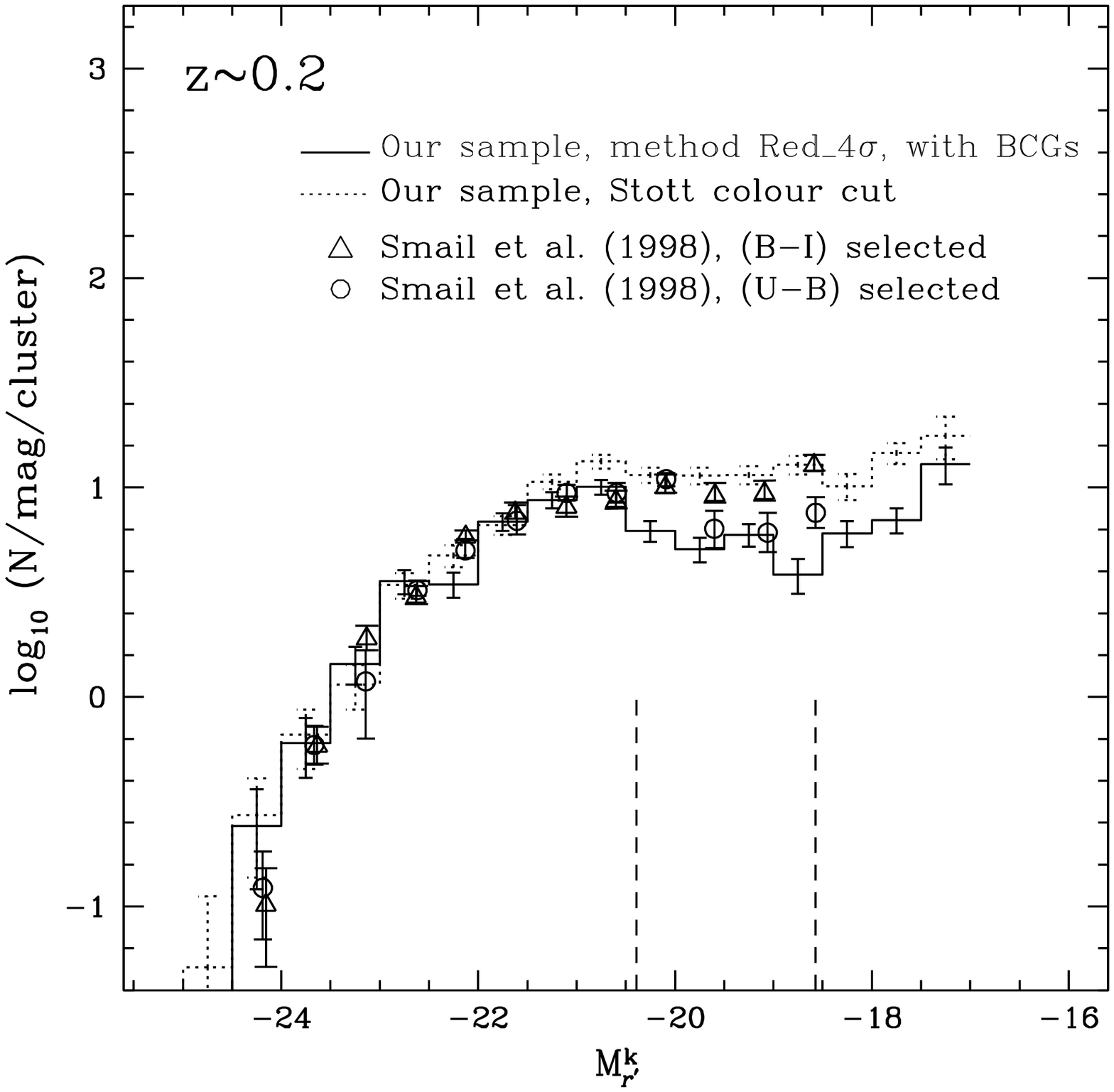}
\caption{Comparison of our red-sequence LFs with those in the literature. Left panel: the solid histogram, filled circles, and open circles represent the red-sequence LFs of the Yang07 sample, \protect \cite{barkhouse} and \protect \cite{stott}, all passively evolved to $z=0.2$. Right panel: the solid histogram is the LF of our sample at $z\sim0.2$, calculated using method Red\_$4\sigma$, with the brightest cluster galaxies included for easy comparison with the study from \protect \cite{smail98}.  Circles and triangles are red-sequence LFs from \protect \cite{smail98}, with red-sequence galaxies selected using $(U-B)$, and $(B-I)$ colour respectively. The dotted histogram is the LF of our sample derived using the colour cut adopted by \protect \cite{stott}. Again, Magnitudes are $k$-corrected to rest-frame at $z=0$. The two vertical dashed lines show the approximate division between dwarf and giant as in \protect \cite{lucia}. See text for discussion.\label{litfit_diff}}
\end{figure*}

In the work by \cite{barkhouse}, the red-sequence galaxies are selected in a way that is very similar to our method Red\_$4\sigma$ (see \S \ref{rel}); therefore it is directly comparable to our measurement of the Yang07 sample (Figure \ref{litfit_z}) given the similar redshift. In the left panel of  Figure \ref{litfit_diff}, the solid histogram is the red-sequence LF of  a sample of Abell clusters that are X-ray detected in the EINSTEIN X-ray images \citep{eist} used in \cite{barkhouse}, measured within the same physical radius ($<0.5$ Mpc) as in this work (Barkhouse, priv. comm.). The filled circles represent the LF we measured for the Yang07 sample. The two LFs agree fairly well with each other in general. Note the LFs are normalized to give the same cumulative number count down to $M_{r'}= -20.5$ (around $m^*+2$); if a slightly fainter magnitude cut is used, given the relatively large uncertainties at the bright end, the two LFs would then have an even better agreement at the faint magnitudes. A simple way to circumvent the issue of the relative normalization is to calculate the DGR. The DGR calculated from the LF of the sample from \cite{barkhouse} is shown as the filled circle in Figure \ref{DGRp5Mpcbus}, and it is consistent with that from the Yang07 sample (the square). At a similar redshift, \cite{stott} studied 10 X-ray-luminous ($L_X>5\times 10^{44}\ ergs\ s^{-1}$) clusters at $\bar{z}=0.13$ from the Las Campanas/AAT Rich Cluster Survey (LARCS; \citealt{larcs01,larcs06}), and their LF is shown as the open circles in Figure \ref{litfit_diff}. It has a steeper faint end than that from both \cite{barkhouse} and the Yang07 sample. We attribute this difference to the way red-sequence galaxies are selected: in the work by \cite{stott}, red-sequence galaxies are considered as those within a fixed colour cut  around the CMR ($|\triangle(B-R)|<0.4$ mag). This cut is very wide, and therefore would include contamination from the blue cluster members, overestimating the number of faint red-sequence galaxies. Other works at a similar redshift includes, for example, \cite{tanaka} and \cite{hansen07}. We do not compare our LFs with that in those works here because, in \cite{tanaka} the cluster environment is defined using local density, and thus is not clear within what physical radius the measurement is made. And in the work by \cite{hansen07} the cluster sample spans over a  redshift range ($0.1<z<0.3$) within which we see evolution in this work; therefore it is not directly comparable.

In the right panel of Figure \ref{litfit_diff}, the solid histogram is the LF of our sample at $z\sim0.2$, calculated using method Red\_$4\sigma$, with the brightest cluster galaxies included for easy comparison with the study from \cite{smail98}.  Circles and triangles are red-sequence LFs of a sample of 10 X-ray clusters from ROSAT All Sky Survey at $z=0.22-0.28$ from \cite{smail98}. Circles represent red galaxies selected using $(U-B)$ colour with a width from $0.28-0.43$ mag as a function of magnitude, and triangles are selected using $(B-I)$ colour with a width from $0.18-0.33$ mag. The filter combination of $(B-I)$ is similar to $(g'-r')$, and the relatively narrow colour slice with varying width as a function of magnitude makes it more similar to our method P10\_$2\sigma$. The LF selected using $(U-B)$ has a better agreement with our LF from method Red\_$4\sigma$, which could be due to the fact that at $z\sim 0.2$ the $(U-B)$ colour is more sensitive to recent star formation than $(B-I)$, and therefore the population that is red in $(U-B)$ is less contaminated by the cluster blue cloud. As a result, the DGR from this LF would be located close to that of our sample at $z\sim 0.2$ if plotted in Figure \ref{DGRp5Mpcbus}. To show how a wide colour cut affects the LF and DGR of our sample, we calculate the LF of our sample using the colour cut of \cite{stott}. We transform the difference in $(B-R)$ into $(g'-r')$ making use of the colour difference between E and Sab galaxies in $(B-R)$ and in $(g'-r')$ at $z=0.2$ from \cite{fuku}. The resulting LF for the sample at $z\sim 0.2$ is shown as the dotted histogram in the right panel of Figure \ref{litfit_diff}, and the DGRs are shown as crosses in  Figure \ref{DGRp5Mpcbus}. Comparing with our LF calculated using method Red\_$4\sigma$, it produces significantly more faint red-sequence galaxies, and as a result gives a higher DGR. This shows that the faint end is sensitive to the selection method; therefore when looking for evolution with the redshift, it is important to make sure all methods are consistent.

The DGR of Coma we calculated ($2.5\pm 0.8$) is comparable to that in other studies in the literature that used statistical background subtraction \citep[e.g.][]{lucia,stott,andreon08}. Also, a complete spectroscopic study on Coma by Marzke (priv. comm.) obtained a red-sequence DGR of $\sim 2.1$, which confirms  the results from statistical studies.

The red-sequence DGRs from the studies in the literature are compiled in the work by \cite{gb} (see their figure 1). In the redshift range  $0.1\lsim z \lsim0.4$, our measurements of the DGRs are systematically lower than those in the literature (but note most of those studies do not directly overlap with this redshift range). Given that the available studies in the literature for the compilation all used different selection criteria for red-sequence galaxies, and we have shown that the faint end and thus the DGR is sensitive to the selection method, we suggest the discrepancy is mostly due to how red-sequence galaxies are selected.

\subsection{The Relation between Red Galaxies and Passive Galaxies}
Our purpose of studying red-sequence galaxies is to understand how their star formation is shut off as they move from the blue cloud; therefore we want to select a population that is as close to a truely quiescent population as possible.

The population selected using colour that brackets the 4000\AA \ break probably is very close to a population that is truely dead, however it can be contaminated by dusty star-forming galaxies \citep[e.g.][]{dusty}. Since FIR (24$\mu m$) emission is sensitive to dusty star formation, it can be used to estimate the contamination from the dusty star-forming galaxies with red colours, and to select passive populations.  The CFHTLS W1 field overlaps with the $Spitzer$ Wide-area Infrared Extragalactic Survey (SWIRE) XMM-LSS field, but the depth of the 24 $\mu m$ data  is not enough to estimate the contamination of the dusty star-forming galaxies down to low enough SFR at the redshifts probed in this work. According to the work by \cite{dusty}, in the core of clusters, the contamination of dusty star-forming galaxies with SFR down to 0.14 $M_{\odot}$ yr$^{-1}$ is $\sim 20$ per cent, dominating at $M_* >10^{10} M_{\odot}$, and $<10$ per cent at $M_* <10^{10} M_{\odot}$. However, the contamination rate is likely to be lower because in that work not enough information is available to eliminate the contribution from AGNs. Therefore, we believe our LF is an  accurate representation of the truly inactive population.

Other than colour, galaxies can be categorized according to their morphology as well. Traditionally, elliptical and lenticular galaxies are termed early-type galaxies, while spiral galaxies are considered late-type galaxies. Also, emission-line strength can be used to broadly divide galaxies into emission-line (star-forming) or non-emission-line (non-star-forming) galaxies. 
These methods may not select exactly the same population. The morphological selection does not always trace star formation: although elliptical galaxies are generally considered as non-star-forming, observations have indicated that a fraction of elliptical galaxies do show recent star formation \citep[e.g.][]{ESF}; thus they would be included in a morphologically selected sample. On the other hand, there could be spiral galaxies that do not have star formation, as indicated by the partially independent transformation in morphology and star formation rate in cluster environments \citep[e.g.][]{Ha}; thus these spiral galaxies would be missed. Therefore, we compare the  passive population selected using these different methods.

\cite{godwin} constructed the LFs of Coma for different morphological types using the magnitude measured from photographic plates. Based on the numbers read off their figure 4, the DGR of the elliptical+lenticular galaxies is $\sim 2.2$. This is consistent with the DGR of the red-sequence galaxies in Coma selected using colour \citep[this work, $2.5\pm 0.8$; Marzke (priv. comm.);][]{lucia,stott,andreon08}.

\cite{OII}  constructed $R-$band LF for six Abell clusters at $0.013\leq z\leq 0.067$, and split them into emission-line and non-emission-line samples, based on their [OII] strength. The DGR we calculate from their LF of the non-emission-line cluster galaxies ($2.0\pm 0.2$), is between that of the Yang07 sample and Coma, consistent with the trend seen in Figure \ref{DGRp5Mpcbus} (although note the criterion of EW[OII]$<$5\AA \ used to select non-emission-line galaxies in their work still allows a moderate level of star formation). These comparsions suggest that at low redshift, colour, morphology or emission-line selects mostly the same passive population. This is consistent with the studies by for example, \cite{SS92,strateva01} and \cite{hogg02}.

\subsection{Possible Systematics}\label{sys}
\subsubsection{Local vs. Global Background}
We have used a global background subtraction to construct the LFs described in \S \ref{rslf}. However, since there are associated large-scale structures around clusters, the background around clusters might be different from the random field. Therefore, we test this effect by constructing the LFs for our sample using global background subtraction. Instead of taking 500 random fields, for each cluster we take the galaxy count in an annulus of radius 10 Mpc around the cluster centre, with an area equivalent to the $r<0.5$ Mpc core, as the background. The resulting LFs are consistent with those shown in Figure \ref{LFz3bin}. Thus, we conclude that the use of a global background does not bias our results. 

Even if our background is underestimated due to the presence of large-scale structure, this will not have a significant effect on the red-sequence DGR. As can be seen from  Figure \ref{LFz3bin}, using  method Red\_$4\sigma$ the background is negligible except at low redshift and at magnitudes fainter than those used to define the dwarf population.

Note that the LFs we derived, and those in the literature with which we compared, are of the projected galaxy distribution. Thus, they include galaxies from $r>0.5$ Mpc. This may lead to a faint end slope which is somewhat steeper than that of the genuine core population, but would not significantly affect the DGR measurement \citep[e.g.][]{barkhouse}.

\subsubsection{Systematic Uncertainties in Photometry}\label{d1}
In this section, we examine any systematic uncertainties in photometry.
We make use of the data in D1 to examine whether the shallower faint end of our LFs is due to incompleteness, or biased photometry. Two clusters in our sample are in both W1 and D1. We construct the LF based on the data from D1 and compare it with that constructed using data from W1. The two LFs are consistent within the uncertainty, although the error bars are rather large, given the small number of clusters. Thus, the lack of low-luminosity galaxies at $0.2\lsim z \lsim 0.4$ we find is not due to insufficient depth or biased photometry. This is not unexpected, as we have shown we are $100$ per cent complete, with small errors at the relevant magnitude, based on comparison with the same D1 field.

\addtocounter{footnote}{-1}
As an external check, we extract from the SDSS archive a subset of galaxies in a 2x2 square degree region that overlaps with the CFHTLS W1 field, and compare the total magnitude (mag\_auto for CFHTLS, and mag\_model for SDSS) of the common galaxies. After transforming the CFHTLS magnitude to SDSS magnitude using the formula provided on the Terapix website$^\dagger$\footnote{$^\dagger$ http://terapix.iap.fr/rubrique.php?id\_rubrique=241}, the median of the distribution of the magnitude difference in  $g'$, $r'$ and $i'$ filters are all less than $\sim 0.05$ mag, and with no obvious trend with magnitude.

Therefore, the uncertainties in the CFHTLS photometry is not a major source of error.

\subsubsection{Redshift Accuracy}\label{smooth}
As described in \S \ref{ppz}, the redshifts we assign to our clusters are accurate to the $\sim 0.01$ level. If the estimated redshift of a cluster is slightly higher or lower than what it actually is, the absolute magnitudes of galaxies in that cluster converted from apparent magnitudes will be brighter or fainter than what they should be, and the intrinsic shape of the LF will be distorted. Here we examine how big this effect is using the Schechter function fit from \cite{barkhouse}. We simulate a sample of 100 clusters at $z=0.2$, each of them with the same intrinsic LF as the Schechter function fit from \cite{barkhouse}. We use a Monte Carlo method, randomly assigning a redshift to each cluster in the range of $z=0.2\pm 0.01$. This is equivalent to randomly shifting the LF horizontally toward brighter or fainter magnitudes by $0.29$ mag. We then take the average of the 100 shifted LFs as the stacked LF for this sample. The stacked LF only makes $\sim$1.3 per cent difference to the bright end (brighter than $M_{r'}=-21$) of the undistorted LF, and $\sim$0.1 per cent to the faint end. Therefore, this does not affect our results. Note, this uncertainty of $\triangle z \sim 0.01$ is comparable to the redshift interval between two adjacent colour slices; thus the effect of projection (\S \ref{proj}) is of comparable uncertainty. 

\subsubsection{Aperture Effect}

As mentioned in \S \ref{phot}, the colour is measured within a fixed aperture of diameter 4.7 arcsec. To examine how colour gradients of galaxies affect our results, we measure the colour in a smaller aperture, of diameter 3 arcsec, and compare the colours measured in the two different apertures. The distribution of the colour difference of the red-sequence galaxies in the two apertures is symmetric in general. The scattering of the distribution is smaller than or comparable to the uncertainty on the colour determined using the overlapping pointings (see Figure \ref{grerrlinfit}). At the bright end, the median of the colour in the larger aperture is $\sim 0.03$ mag bluer (smaller than the width of the colour slice).  For the LFs constructed using methods P10\_$2\sigma$ and NP\_$2\sigma$, the number of red-sequence galaxies might be slightly underestimated due to this scattering. However, for methods Red\_$4\sigma$ and Red\_all, the results are not affected as long as the scattering is symmetric.

\subsubsection{Richness Dependence}\label{richdgr}
Since we detect clusters using only galaxies brighter than $m^*$+2, which is around the magnitude that separates giant from dwarf galaxies, it is possible that our sample is biased towards systems with more giants. To test this, we examine whether clusters with fewer giants have higher DGRs. We calculate the DGRs for a subset of very rich clusters with $N_{red,m^*+2}\geq 18$ (plotted as open pentagons in Figure \ref{DGRp5Mpcbus}, with slight offset in redshift for clarity), and for a set of clusters that are not rich enough to be included in our final catalogue ($7 \leq N_{red,m^*+2}\leq 10$), using our best method Red\_$4\sigma$. We do not find strong dependence of the DGR on the richness of the clusters. Therefore, our results are not biased due to the detection method. Note the richness of our clusters is measured within a radius of 0.5 Mpc from cluster centres, and all analysis are carried out for the same region. To check if this biases our results, we calculate the DGR using a slightly larger radius  ($r<$1 Mpc), and find that the change of DGR is within $\sim 1 \sigma$ of the values shown by the triangles in Figure \ref{DGRp5Mpcbus}; thus it does not change our conclusions either. 

For clusters with $N_{red,m^*+2}\geq 12$, we do not detect obvious correlation between the individual DGR and  $N_{red,m^*+2}$. For the richest cluster in our sample, Abell0362 at $z\sim 0.2$ (\S \ref{pprich}), its red-sequence DGR is $\sim 0.7\pm 0.2$, similar to the sample average DGR. The deficit of dwarf galaxies ($20.5<r'<22.3$) is clearly visible on its CMD in Figure \ref{mostmassive}. For the four clusters in our sample that have comparable richness as Coma, their average DGR is $0.9 \pm 0.1$, similar to the sample average as well. We also constructed the composite  red-sequence LF for the four X-ray confirmed clusters that have $N_{red,m^*+2}\geq 12$ (but they are all poor clusters with kT$\sim 2$ keV), and it is consistent with the LF of the whole sample, within the uncertainties. 

Note although Coma is $\sim 4$ times richer than the typical clusters in our sample and the Yang07 sample, the lack of strong richness dependence of the red-sequence DGR seen in our sample suggests that the rapid evolution between $z\sim 0.2$ and $z\sim 0$ we see in Figure \ref{DGRp5Mpcbus} is probably a genuine redshift evolution. However, this is based on the assumption that the lack of richness dependence still holds at $z\sim 0$, and that Coma is a typical cluster at this redshift, that can be fairly compared with our ensemble averages. We will use a larger low redshift comparison sample to explore this issue in a future paper.

Note, at higher redshifts, $z\sim 0.5$, the study by \cite{gilbanklf}, using a statistical sample of red-sequence selected clusters (similar to that presented in this work), found that the red-sequence DGR is higher for richer systems  than that for poorer systems. This is contrary to the findings of \cite{lucia}, who used $\sim 10$ optically selected clusters at $z\sim 0.4-0.8$. The richness cuts used in  \cite{gilbanklf} are $B_{gc}>800$ for the richer subsample (the richness parameter, $B_{gc}$, is described in \citealt{bgc}), and $300<B_{gc}<500$ for the poorer subsample. \cite{lucia} calculated the dynamical masses (rather than richnesses) via velocity dispersions for individual clusters. The criterion they used to separate ``richer'' (higher velocity dispersion) and ``poorer'' (lower velocity dispersion) clusters was the velocity dispersion being higher or lower than $600 $ km s$^{-1}$ (which is approximately $B_{gc}\sim 600$, \citealt{gilbanklf}). The division used in both studies roughly corresponds to $N_{red,m^*+2}\sim 12$ in our cluster sample, since the virial mass of a cluster with  $\sigma= 600 $ km s$^{-1}$ and $r_{200}\sim 1$ Mpc is  $\sim 1\times 10^{14} M_{\odot}$. Thus the richness/mass-dependence of the red-sequence DGR at $z\gsim 0.4$ is still an open question.

\subsubsection{Projection Effect}\label{proj}
If two clusters are close enough in redshift space (within neighbouring colour slices), and are projected along the line-of-sight, their red-sequences would not be distinguished.  As a  consequence of this projection effect, the number count of galaxies per cluster  would  be overestimated, and the shape of the LFs would be slightly smoothed (although this latter effect is small, see \S \ref{smooth}). Another consequence is that it might reduce any trend of the DGR with richness, since a rich system that consists of two projected poorer systems would give a DGR that is similar to the poorer systems. 

According to, for example, \cite{gilbanklf}, for clusters in  the Red-Sequence Cluster Survey \citep{rcs} over the redshift range $0.4<z<1$, the contamination rate from projected systems is $\sim 5-10$ per cent. Since the method we use to detect our clusters is based on the CRS method, we expect the contamination rate to be similar. In a later paper, we will use the full CFHTLS data to explore this more thoroughly.

\subsubsection{Slope of the CMR}
Since our method Red\_$4\sigma$ mirrors galaxies on the redder side of the best-fit CMR to avoid contamination from the cluster blue cloud, it is important that the best-fit CMR goes through the centre of the red-sequence. Because of the heavy contamination from the blue cloud at faint magnitudes, the fit to the CMR is carried out at brighter magnitudes down to $m^*+2$, and extrapolated to fainter magnitudes. To test the effect of the fitted slope on the DGR, we fix the colour of the best-fit CMR at $m^*$, vary the slope by $\pm \sim 20$ per cent, and recalculate the red-sequence DGR using method Red\_$4\sigma$. The variation of the DGR due to this slope change is about 20 per cent, and thus does not change our conclusions.

The extrapolation of CMR from bright magnitudes to fainter ones implicitly assumes that the slope is independent of magnitude. This is reasonable for a truely passive population that is already dead, which is what we are after here.  For galaxies that are in the transition of moving from the blue cloud onto the faint end of the red-sequence, we would expect the colour of the red-sequence galaxies to be bluer towards fainter magnitudes; therefore a constant slope over the whole magnitude range would underestimate this population in transition. Which method to use depends on which population one is after. 

\subsubsection{Filter Transformation}
Since the division between dwarf and giant galaxy is defined in the $z=0$ rest-frame $V$ band in the Vega system \citep{lucia}, we now examine the effect of the uncertainty in passband transformation on the red-sequence DGR.

As mentioned in \S \ref{mod}, we use a simple single-burst model, calibrated with the data of Coma, to produce the red-sequence CMD. The $(V-r')$ colour we use to transform the definition of dwarf and giant in $V$ to $r'$ is produced by the same model. The uncertainty associated with this transformation could come from a few sources: for example, the magnitude-metallicity relation fitted to reproduce the CMR of Coma, which, if slightly different, would produce a different colour for a given magnitude; the uncertainties on the photometry of the Coma data \citep{bower92} could also introduce an error on the $(V-r')$ colour. However, given that the spectral energy distribution of the old galaxy population is well known, and over the redshift range studied in this work the observed $r'$ band is similar to the rest-frame $V$ band, we expect the uncertainty on $(V-r')$ to be small, at a level of $<0.1$ mag.

We now examine how an uncertainty of $\pm 0.1$ mag on the division between dwarf and giant in $r'$ would affect our results. To maximize the effect, we make the dividing point between dwarf and giant 0.1 mag brighter, and the faint limit of dwarf 0.1 mag fainter. The resulting DGRs for our cluster sample calculated exactly the same way as in  \S \ref{ourcl} are $0.8\pm 0.05$ at $z\sim 0.2$; $1.0\pm 0.05$ at $z\sim 0.27$, and $1.1\pm 0.06$ at $z\sim 0.33$. Note this demonstrates the maximum effect by taking an extreme case (two offsets of 0.1 mag in directions chosen to maximise changes in DGR); the real impact on the results due to the filter transformation is smaller than this.
\\

None of the systematics discussed in this section significantly affects our results. Thus, the lack of evolution at $0.2\lsim z\lsim 0.4$, and the stronger evolution since $z\sim 0.2$ seen is robust.

\section{Conclusions}\label{con}

In this work we detected and compiled a large sample of 127 clusters over the redshift range of $0.17\leq z \leq 0.36$ from the CFHTLS data using a method similar to the CRS method \citep{GY00}, and constructed the $r'-$band red-sequence LFs within a projected radius of $r<0.5$ Mpc from cluster centres. The depth of the CFHTLS data enables us to go deeper at these redshifts for a larger sample than previous studies. Our main results are as follows:\\
$\bullet$ We have presented a thorough study of the effect of colour selections on the red-sequence LF, and showed that the faint end of the LF is very sensitive to how red-sequence galaxies are selected. We find that one optimal way to minimise the blue cloud contamination is to mirror galaxies redder (but no redder than $4\sigma$) than the CMR, with a second colour cut to further reduce the background.\\
$\bullet$ The red-sequence LFs of our sample have a significant inflexion centred at $M_{r'}\sim -18.5$ (especially at $z\sim 0.2$), and thus cannot be described by a single Schechter function, suggesting a mixture of two populations. \\
$\bullet$ By comparing the red-sequence LFs of our sample with that of low redshift samples constructed from SDSS, calculated consistently using our best method, we showed there is a steepening of the faint end with decreasing redshift since $z\sim 0.2$, but no strong evolution between $z\sim 0.2$ and $z\sim 0.4$. \\
$\bullet$ As a result of the evolution of the LFs, the red-sequence DGRs we measured show no significant changes over $0.2\lsim z \lsim 0.4$, but an increase of a factor of $\sim 3$ from $z\sim 0.2$ to $z\sim 0$ (over the last 2.5 Gyr). Also, we do not see a strong  dependence of the DGR on the richness of the clusters within our own sample.\\
$\bullet$ We thoroughly checked possible systematics, and showed none of them significantly affects our results, and thus our results are robust.\\

The lack of evolution of the red-sequence DGR over $0.2\lsim z \lsim 0.4$, and the increase since $z\sim 0.2$, suggest a rapid build-up of the cluster red-sequence in terms of the {\it number} of dwarf galaxies over the last 2.5 Gyr (the corresponding {\it stellar mass} on the red-sequence has only increased by 15 per cent). Taking the time delay assumed in \cite{redfrac} for galaxies to move onto the red-sequence (1 Gyr for the single burst SSP models, and 0.5 Gyr for the quenched models), our results imply a significant fraction of faint red galaxies have been quenched within the last $\sim 3$ Gyr.

At higher redshifts than probed in this work, the evolution of the faint end of the red-sequence is still controversial, see for example \cite{stott}, \cite{lucia}, and \cite{andreon08}. Given the different colour selections employed in those studies, and the sensitivity of the faint end to the selection method we have demonstrated, in a subsequent paper, we will extend our current study and measure the red-sequence LFs consistently out to higher redshifts. With the release of new data from CFHTLS, we will have a larger cluster sample, and thus better statistics.

\section*{Acknowledgment}
We thank the referee for useful comments. We also thank Stefano Andreon, Mike Hudson, Steven Allanson and Sean McGee for helpful discussions, and Wayne Barkhouse and Ron Marzke for kindly providing their data.

 This work was supported by an Early Researcher Award from the province of Ontario and an NSERC Discovery Grant to MLB, and was made possible by the facilities of the Shared Hierarchical Academic Research Computing Network (SHARCNET:www.sharcnet.ca).

Based on observations obtained with MegaPrime/MegaCam, a joint project of CFHT and CEA/DAPNIA, at the  Canada-France-Hawaii Telescope (CFHT) which is operated by the National Research Council (NRC) of Canada, the Institut National des Science de l'Univers of the Centre National de la Recherche Scientifique (CNRS) of France, and the University of Hawaii. This work is based in part on data products produced at TERAPIX and the Canadian Astronomy Data Centre as part of the Canada-France-Hawaii Telescope Legacy Survey, a collaborative project of NRC and
CNRS.

 Funding for the SDSS and SDSS-II has been provided by the Alfred P. Sloan Foundation, the Participating Institutions, the National Science Foundation, the U.S. Department of Energy, the National Aeronautics and Space Administration, the Japanese Monbukagakusho, the Max Planck Society, and the Higher Education Funding Council for England. The SDSS Web Site is http://www.sdss.org/. The SDSS is managed by the Astrophysical Research Consortium for the Participating Institutions. The Participating Institutions are the American Museum of Natural History, Astrophysical Institute Potsdam, University of Basel, University of Cambridge, Case Western Reserve University, University of Chicago, Drexel University, Fermilab, the Institute for Advanced Study, the Japan Participation Group, Johns Hopkins University, the Joint Institute for Nuclear Astrophysics, the Kavli Institute for Particle Astrophysics and Cosmology, the Korean Scientist Group, the Chinese Academy of Sciences (LAMOST), Los Alamos National Laboratory, the Max-Planck-Institute for Astronomy (MPIA), the Max-Planck-Institute for Astrophysics (MPA), New Mexico State University, Ohio State University, University of Pittsburgh, University of Portsmouth, Princeton University, the United States Naval Observatory, and the University of Washington. 

The simulations in this paper were carried out by the Virgo Supercomputing Consortium using computer s based at the Computing Centre of the Max-Planck Society in Garching and at the Edinburgh parallel Computing Centre. The data are publicly available at http://www.mpa-garching.mpg.de/NumCos.


\begin{thebibliography}{52}
\expandafter\ifx\csname natexlab\endcsname\relax\def\natexlab#1{#1}\fi

\bibitem[\protect\citeauthoryear{{Allanson}, {Hudson}, {Smith} \&
  {Lucey}}{{Allanson} et~al.}{2009}]{steve}
{Allanson} S.,  {Hudson} M.,  {Smith} R.,    {Lucey} J.,  2008, ApJ, submitted

\bibitem[\protect\citeauthoryear{{Andreon}}{{Andreon}}{2006}]{andreon}
{Andreon} S.,  2006, MNRAS, 369, 969

\bibitem[\protect\citeauthoryear{{Andreon}}{{Andreon}}{2008}]{andreon08}
{Andreon} S.,  2008, MNRAS, 386, 1045


\bibitem[\protect\citeauthoryear{{Baldry}, {Balogh}, {Bower}, {Glazebrook},
  {Nichol}, {Bamford} \& {Budavari}}{{Baldry} et~al.}{2006}]{baldry06}
{Baldry} I.~K.,  {Balogh} M.~L.,  {Bower} R.~G.,  {Glazebrook} K.,  {Nichol}
  R.~C.,  {Bamford} S.~P.,    {Budavari} T.,  2006, MNRAS, 373, 469

\bibitem[\protect\citeauthoryear{{Baldry}, {Glazebrook}, {Brinkmann},
  {Ivezi{\'c}}, {Lupton}, {Nichol} \& {Szalay}}{{Baldry} et~al.}{2004}]{baldry}
{Baldry} I.~K.,  {Glazebrook} K.,  {Brinkmann} J.,  {Ivezi{\'c}} {\v Z}.,
  {Lupton} R.~H.,  {Nichol} R.~C.,    {Szalay} A.~S.,  2004, ApJ, 600, 681

\bibitem[\protect\citeauthoryear{{Balogh}, {Eke}, {Miller}, {Lewis}, {Bower},
  {Couch}, {Nichol}, {Bland-Hawthorn}, {Baldry}, {Baugh}, {Bridges}, {Cannon},
  {Cole}, {Colless}, {Collins}, {Cross}, {Dalton}, {de Propris}, {Driver},
  {Efstathiou}, {Ellis}, {Frenk}, {Glazebrook}, {Gomez},
  {Gray}, {Hawkins}, {Jackson}, {Lahav}, {Lumsden}, {Maddox}, {Madgwick},
  {Norberg}, {Peacock}, {Percival}, {Peterson}, {Sutherland} \&  {Taylor}}{{Balogh} et~al.}{2004{\natexlab{a}}}]{balogh04b}
  {Balogh} M., et al. \  2004{\natexlab{a}}, MNRAS, 348, 1355

\bibitem[\protect\citeauthoryear{{Balogh}, {Baldry}, {Nichol}, {Miller},
  {Bower} \& {Glazebrook}}{{Balogh} et~al.}{2004{\natexlab{b}}}]{balogh04}
{Balogh} M.~L.,  {Baldry} I.~K.,  {Nichol} R.,  {Miller} C.,  {Bower} R.,
  {Glazebrook} K.,  2004{\natexlab{b}}, ApJL, 615, L101

\bibitem[\protect\citeauthoryear{{Barkhouse}, {Yee} \&
  {L{\'o}pez-Cruz}}{{Barkhouse} et~al.}{2007}]{barkhouse}
{Barkhouse} W.~A.,  {Yee} H.~K.~C.,    {L{\'o}pez-Cruz} O.,  2007, ApJ, 671,
  1471

\bibitem[\protect\citeauthoryear{{Bell}, {Wolf}, {Meisenheimer}, {Rix},
  {Borch}, {Dye}, {Kleinheinrich}, {Wisotzki} \& {McIntosh}}{{Bell}
  et~al.}{2004}]{bell}
{Bell} E.~F.,  et al.  2004, ApJ,
  608, 752

\bibitem[\protect\citeauthoryear{{Bertin} \& {Arnouts}}{{Bertin} \&
  {Arnouts}}{1996}]{sextractor}
{Bertin} E.,  {Arnouts} S.,  1996, A\&AS, 117, 393

\bibitem[\protect\citeauthoryear{{Binggeli}, {Sandage} \& {Tammann}}{{Binggeli}
  et~al.}{1988}]{binggeli}
{Binggeli} B.,  {Sandage} A.,    {Tammann} G.~A.,  1988, ARA\&A, 26, 509




\bibitem[\protect\citeauthoryear{{Blanton}, {Dalcanton}, {Eisenstein},
  {Loveday}, {Strauss}, {SubbaRao}, {Weinberg}, {Anderson}, {Annis}, {Bahcall},
  {Bernardi}, {Brinkmann}, {Brunner}, {Burles}, {Carey}, {Castander},
  {Connolly}, {Csabai}, {Doi}, {Finkbeiner}, {Friedman}, {Frieman}, {Fukugita},
  {Gunn}, {Hennessy}, {Hindsley}, {Hogg}, {Ichikawa}, {Ivezi{\'c}}, {Kent},
  {Knapp}, {Lamb}, {Leger}, {Long}, {Lupton}, {McKay}, {Meiksin}, {Merelli},
  {Munn}, {Narayanan}, {Newcomb}, {Nichol}, {Okamura}, {Owen}, {Pier}, {Pope},
  {Postman}, {Quinn}, {Rockosi}, {Schlegel}, {Schneider}, {Shimasaku},
  {Siegmund}, {Smee}, {Snir}, {Stoughton}, {Stubbs}, {Szalay}, {Szokoly},
  {Thakar}, {Tremonti}, {Tucker}, {Uomoto}, {Vanden Berk}, {Vogeley},
  {Waddell}, {Yanny}, {Yasuda}, \& {York}}{{Blanton} et~al.}{2001}]{Blanton01}
{Blanton} M.~R.,  et al.\ 2001, AJ, 121, 2358

\bibitem[\protect\citeauthoryear{{Bower}, {Benson}, {Malbon}, {Helly}, {Frenk},
  {Baugh}, {Cole} \& {Lacey}}{{Bower} et~al.}{2006}]{bowermodel}
{Bower} R.~G.,  {Benson} A.~J.,  {Malbon} R.,  {Helly} J.~C.,  {Frenk} C.~S.,
  {Baugh} C.~M.,  {Cole} S.,    {Lacey} C.~G.,  2006, MNRAS, 370, 645

\bibitem[\protect\citeauthoryear{{Bower}, {Lucey} \& {Ellis}}{{Bower}
  et~al.}{1992}]{bower92}
{Bower} R.~G.,  {Lucey} J.~R.,    {Ellis} R.~S.,  1992, MNRAS, 254, 601

\bibitem[\protect\citeauthoryear{{Bruzual} \& {Charlot}}{{Bruzual} \&
  {Charlot}}{2003}]{galexev}
{Bruzual} G.,  {Charlot} S.,  2003, MNRAS, 344, 1000

\bibitem[\protect\citeauthoryear{{Christlein} \& {Zabludoff}}{{Christlein} \&
  {Zabludoff}}{2003}]{OII}
{Christlein} D.,  {Zabludoff} A.~I.,  2003, ApJ, 591, 764

\bibitem[\protect\citeauthoryear{{Cimatti}, {Daddi} \& {Renzini}}{{Cimatti}
  et~al.}{2006}]{cimatti06}
{Cimatti} A.,  {Daddi} E.,    {Renzini} A.,  2006, A\&A, 453, L29

\bibitem[\protect\citeauthoryear{{Colberg}, {White}, {Yoshida}, {MacFarland},
  {Jenkins}, {Frenk}, {Pearce}, {Evrard}, {Couchman}, {Efstathiou}, {Peacock},
  {Thomas} \& {The Virgo Consortium}}{{Colberg} et~al.}{2000}]{colberg}
{Colberg} J.~M., et al.  2000, MNRAS, 319, 209

\bibitem[\protect\citeauthoryear{{Crawford}, {Bershady} \&
  {Hoessel}}{{Crawford} et~al.}{2009}]{crawford}
{Crawford} S.~M.,  {Bershady} M.~A.,    {Hoessel} J.~G.,  2009, ApJ, 690, 1158

\bibitem[\protect\citeauthoryear{{Cruddace}, {Voges}, {B{\"o}hringer},
  {Collins}, {Romer}, {MacGillivray}, {Yentis}, {Schuecker}, {Ebeling} \& {De
  Grandi}}{{Cruddace} et~al.}{2002}]{abell0362}
{Cruddace} R., et al.  2002, ApJS, 140, 239

\bibitem[\protect\citeauthoryear{{De Lucia}, {Poggianti},
  {Arag{\'o}n-Salamanca}, {White}, {Zaritsky}, {Clowe}, {Halliday}, {Jablonka},
  {von der Linden}, {Milvang-Jensen}, {Pell{\'o}}, {Rudnick}, {Saglia} \&
  {Simard}}{{De Lucia} et~al.}{2007}]{lucia}
{De Lucia} G.,  et al.\ 2007, MNRAS, 374, 809

\bibitem[\protect\citeauthoryear{{Evrard}, {MacFarland}, {Couchman}, {Colberg},
  {Yoshida}, {White}, {Jenkins}, {Frenk}, {Pearce}, {Peacock} \&
  {Thomas}}{{Evrard} et~al.}{2002}]{hubble}
{Evrard} A.~E.,  et al.\ 2002, ApJ, 573, 7

\bibitem[\protect\citeauthoryear{{Evrard}, {Metzler} \& {Navarro}}{{Evrard}
  et~al.}{1996}]{evrard96}
{Evrard} A.~E.,  {Metzler} C.~A.,    {Navarro} J.~F.,  1996, ApJ, 469, 494

\bibitem[\protect\citeauthoryear{{Faber}, {Willmer}, {Wolf}, {Koo}, {Weiner},
  {Newman}, {Im}, {Coil}, {Conroy}, {Cooper}, {Davis}, {Finkbeiner}, {Gerke},
  {Gebhardt}, {Groth}, {Guhathakurta}, {Harker}, {Kaiser}, {Kassin},
  {Kleinheinrich}, {Konidaris}, {Kron}, {Lin}, {Luppino}, {Madgwick},
  {Meisenheimer}, {Noeske}, {Phillips}, {Sarajedini}, {Schiavon}, {Simard},
  {Szalay}, {Vogt}, \& {Yan}}{{Faber} et~al.}{2007}]{faber07}
{Faber} S.~M., et al.\ 2007, ApJ, 665, 265

\bibitem[\protect\citeauthoryear{{Fukugita}, {Shimasaku} \&
  {Ichikawa}}{{Fukugita} et~al.}{1995}]{fuku}
{Fukugita} M.,  {Shimasaku} K.,    {Ichikawa} T.,  1995, PASP, 107, 945

\bibitem[\protect\citeauthoryear{{Gilbank} \& {Balogh}}{{Gilbank} \&
  {Balogh}}{2008}]{gb}
{Gilbank} D.~G.,  {Balogh} M.~L.,  2008, MNRAS, 385, L116

\bibitem[\protect\citeauthoryear{{Gilbank}, {Yee}, {Ellingson}, {Gladders},
  {Loh}, {Barrientos} \& {Barkhouse}}{{Gilbank} et~al.}{2008}]{gilbanklf}
{Gilbank} D.~G.,  {Yee} H.~K.~C.,  {Ellingson} E.,  {Gladders} M.~D.,  {Loh}
  Y.-S.,  {Barrientos} L.~F.,    {Barkhouse} W.~A.,  2008, ApJ, 673, 742

\bibitem[\protect\citeauthoryear{{Gladders} \& {Yee}}{{Gladders} \&
  {Yee}}{2000}]{GY00}
{Gladders} M.~D.,  {Yee} H.~K.~C.,  2000, AJ, 120, 2148

\bibitem[\protect\citeauthoryear{{Gladders} \& {Yee}}{{Gladders} \&
  {Yee}}{2005}]{rcs}
{Gladders} M.~D.,  {Yee} H.~K.~C.,  2005, ApJS, 157, 1

\bibitem[\protect\citeauthoryear{{Godwin} \& {Peach}}{{Godwin} \&
  {Peach}}{1977}]{godwin}
{Godwin} J.~G.,  {Peach} J.~V.,  1977, MNRAS, 181, 323

\bibitem[\protect\citeauthoryear{{Gonzalez}, {Zaritsky} \&
  {Wechsler}}{{Gonzalez} et~al.}{2002}]{G+02}
{Gonzalez} A.~H.,  {Zaritsky} D.,    {Wechsler} R.~H.,  2002, ApJ, 571, 129

\bibitem[\protect\citeauthoryear{{Hansen}, {Sheldon}, {Wechsler} \&
  {Koester}}{{Hansen} et~al.}{2007}]{hansen07}
{Hansen} S.~M.,  {Sheldon} E.~S.,  {Wechsler} R.~H.,    {Koester} B.~P.,  2007,
  ArXiv e-prints

\bibitem[\protect\citeauthoryear{{Hogg}, {Blanton}, {Strateva}, {Bahcall},
  {Brinkmann}, {Csabai}, {Doi}, {Fukugita}, {Hennessy}, {Ivezi{\'c}}, {Knapp},
  {Lamb}, {Lupton}, {Munn}, {Nichol}, {Schlegel}, {Schneider} \& {York}}{{Hogg}
  et~al.}{2002}]{hogg02}
{Hogg} D.~W., et al. 2002, AJ, 124,
  646

\bibitem[\protect\citeauthoryear{{Johnston}, {Sheldon}, {Wechsler}, {Rozo},
  {Koester}, {Frieman}, {McKay}, {Evrard}, {Becker} \& {Annis}}{{Johnston}
  et~al.}{2007}]{johnston}
{Johnston} D.~E.,  {Sheldon} E.~S.,  {Wechsler} R.~H.,  {Rozo} E.,  {Koester}
  B.~P.,  {Frieman} J.~A.,  {McKay} T.~A.,  {Evrard} A.~E.,  {Becker} M.~R.,
  {Annis} J.,  2007, ArXiv e-prints

\bibitem[\protect\citeauthoryear{{Jones} \& {Forman}}{{Jones} \&
  {Forman}}{1999}]{eist}
{Jones} C.,  {Forman} W.,  1999, ApJ, 511, 65

\bibitem[\protect\citeauthoryear{{Kaviraj}, {Schawinski}, {Devriendt},
  {Ferreras}, {Khochfar}, {Yoon}, {Yi}, {Deharveng}, {Boselli}, {Barlow},
  {Conrow}, {Forster}, {Friedman}, {Martin}, {Morrissey}, {Neff},
  {Schiminovich}, {Seibert}, {Small}, {Wyder}, {Bianchi}, {Donas}, {Heckman},
  {Lee}, {Madore}, {Milliard}, {Rich}, \& {Szalay}}{{Kaviraj} et~al.}{2007}]{ESF}
{Kaviraj} S., et al.\ 2007,
  ApJS, 173, 619

\bibitem[\protect\citeauthoryear{{Kodama} \& {Arimoto}}{{Kodama} \&
  {Arimoto}}{1997}]{met_mag}
{Kodama} T.,  {Arimoto} N.,  1997, A\&A, 320, 41

\bibitem[\protect\citeauthoryear{{Kodama}, {Balogh}, {Smail}, {Bower} \&
  {Nakata}}{{Kodama} et~al.}{2004}]{Ha}
{Kodama} T.,  {Balogh} M.~L.,  {Smail} I.,  {Bower} R.~G.,    {Nakata} F.,
  2004, MNRAS, 354, 1103


\bibitem[\protect\citeauthoryear{{Koester}, {McKay}, {Annis}, {Wechsler},
  {Evrard}, {Bleem}, {Becker}, {Johnston}, {Sheldon}, {Nichol}, {Miller},
  {Scranton}, {Bahcall}, {Barentine}, {Brewington}, {Brinkmann}, {Harvanek},
  {Kleinman}, {Krzesinski}, {Long}, {Nitta}, {Schneider}, {Sneddin}, {Voges},
  \& {York}}{{Koester} et~al.}{2007}]{MaxBCG}
{Koester} B.~P., et al.\ 2007, ApJ, 660, 239

\bibitem[\protect\citeauthoryear{{Landy} \& {Szalay}}{{Landy} \&
  {Szalay}}{1993}]{ls93}
{Landy} S.~D.,  {Szalay} A.~S.,  1993, ApJ, 412, 64

\bibitem[\protect\citeauthoryear{{Mercurio}, {Merluzzi}, {Haines}, {Gargiulo},
  {Krusanova}, {Busarello}, {Barbera}, {Capaccioli} \& {Covone}}{{Mercurio}
  et~al.}{2006}]{mercurio06}
{Mercurio} A., et al. 
  2006, MNRAS, 368, 109

\bibitem[\protect\citeauthoryear{{Nelan}, {Smith}, {Hudson}, {Wegner}, {Lucey},
  {Moore}, {Quinney} \& {Suntzeff}}{{Nelan} et~al.}{2005}]{nelan}
{Nelan} J.~E.,  {Smith} R.~J.,  {Hudson} M.~J.,  {Wegner} G.~A.,  {Lucey}
  J.~R.,  {Moore} S.~A.~W.,  {Quinney} S.~J.,    {Suntzeff} N.~B.,  2005, ApJ,
  632, 137




\bibitem[\protect\citeauthoryear{{Pacaud}, {Pierre}, {Adami}, {Altieri},
  {Andreon}, {Chiappetti}, {Detal}, {Duc}, {Galaz}, {Gueguen}, {Le F{\`e}vre}, {Hertling},{Libbrecht},{Melin},{Ponman},{Quintana},{Refregier},{Sprimont},{Surdej},{Valtchanov},{Willis},{Alloin},{Birkinshaw},{Bremer},{Garcet},{Jean}, {Jones},{Le F{\`e}vre},{Maccagni},{Mazure},{Proust},\&  {Trinchieri}}{{Pacaud}
  et~al.}{2007}]{pacaud}
{Pacaud} F.,  et al. \ 
  2007, MNRAS, 382, 1289

\bibitem[\protect\citeauthoryear{{Pimbblet}, {Smail}, {Edge}, {Couch}, {O'Hely}
  \& {Zabludoff}}{{Pimbblet} et~al.}{2001}]{larcs01}
{Pimbblet} K.~A.,  {Smail} I.,  {Edge} A.~C.,  {Couch} W.~J.,  {O'Hely} E.,
  {Zabludoff} A.~I.,  2001, MNRAS, 327, 588

\bibitem[\protect\citeauthoryear{{Pimbblet}, {Smail}, {Edge}, {O'Hely}, {Couch}
  \& {Zabludoff}}{{Pimbblet} et~al.}{2006}]{larcs06}
{Pimbblet} K.~A.,  {Smail} I.,  {Edge} A.~C.,  {O'Hely} E.,  {Couch} W.~J.,
  {Zabludoff} A.~I.,  2006, VizieR Online Data Catalog, 736, 60645

\bibitem[\protect\citeauthoryear{{Popesso}, {Biviano}, {B{\"o}hringer} \&
  {Romaniello}}{{Popesso} et~al.}{2006}]{popesso}
{Popesso} P.,  {Biviano} A.,  {B{\"o}hringer} H.,    {Romaniello} M.,  2006,
  A\&A, 445, 29

\bibitem[\protect\citeauthoryear{{Scarlata}, {Carollo}, {Lilly}, {Feldmann},
  {Kampczyk}, {Renzini}, {Cimatti}, {Halliday}, {Daddi}, {Sargent},
  {Koekemoer}, {Scoville}, {Kneib}, {Leauthaud}, {Massey}, {Rhodes}, {Tasca},
  {Capak}, {McCracken}, {Mobasher}, {Taniguchi}, {Thompson}, {Ajiki}, {Aussel},
  {Murayama}, {Sanders}, {Sasaki}, {Shioya}, \& {Takahashi}}{{Scarlata} et~al.}{2007}]{scarlata}
{Scarlata} C., et al.\ 2007, ApJS, 172, 494

\bibitem[\protect\citeauthoryear{{Schlegel}, {Finkbeiner} \&
  {Davis}}{{Schlegel} et~al.}{1998}]{ext}
{Schlegel} D.~J.,  {Finkbeiner} D.~P.,    {Davis} M.,  1998, ApJ, 500, 525

\bibitem[\protect\citeauthoryear{{Schweizer} \& {Seitzer}}{{Schweizer} \&
  {Seitzer}}{1992}]{SS92}
{Schweizer} F.,  {Seitzer} P.,  1992, AJ, 104, 1039

\bibitem[\protect\citeauthoryear{{Secker} \& {Harris}}{{Secker} \&
  {Harris}}{1996}]{sh96}
{Secker} J.,  {Harris} W.~E.,  1996, ApJ, 469, 623

\bibitem[\protect\citeauthoryear{{Shimizu}, {Kitayama}, {Sasaki} \&
  {Suto}}{{Shimizu} et~al.}{2003}]{T_mass}
{Shimizu} M.,  {Kitayama} T.,  {Sasaki} S.,    {Suto} Y.,  2003, ApJ, 590, 197

\bibitem[\protect\citeauthoryear{{Smail}, {Edge}, {Ellis} \&
  {Blandford}}{{Smail} et~al.}{1998}]{smail98}
{Smail} I.,  {Edge} A.~C.,  {Ellis} R.~S.,    {Blandford} R.~D.,  1998, MNRAS,
  293, 124

\bibitem[\protect\citeauthoryear{{Smith}, {Hudson}, {Lucey}, {Nelan} \&
  {Wegner}}{{Smith} et~al.}{2006}]{SH}
{Smith} R.~J.,  {Hudson} M.~J.,  {Lucey} J.~R.,  {Nelan} J.~E.,    {Wegner}
  G.~A.,  2006, MNRAS, 369, 1419

\bibitem[\protect\citeauthoryear{{Smith}, {Lucey}, {Hudson}, {Allanson},
  {Bridges}, {Hornschemeier}, {Marzke} \& {Miller}}{{Smith}
  et~al.}{2009}]{redfrac}
{Smith} R.~J.,  {Lucey} J.~R.,  {Hudson} M.~J.,  {Allanson} S.~P.,  {Bridges}
  T.~J.,  {Hornschemeier} A.~E.,  {Marzke} R.~O.,    {Miller} N.~A.,  2009,
  MNRAS, 392, 1265

\bibitem[\protect\citeauthoryear{{Springel}, {White}, {Jenkins}, {Frenk},
  {Yoshida}, {Gao}, {Navarro}, {Thacker}, {Croton}, {Helly}, {Peacock}, {Cole},
  {Thomas}, {Couchman}, {Evrard}, {Colberg} \& {Pearce}}{{Springel}
  et~al.}{2005}]{ms}
{Springel} V., et al.\ 2005, 435, 629

\bibitem[\protect\citeauthoryear{{Stott}, {Smail}, {Edge}, {Ebeling}, {Smith},
  {Kneib} \& {Pimbblet}}{{Stott} et~al.}{2007}]{stott}
{Stott} J.~P.,  {Smail} I.,  {Edge} A.~C.,  {Ebeling} H.,  {Smith} G.~P.,
  {Kneib} J.-P.,    {Pimbblet} K.~A.,  2007, ApJ, 661, 95

\bibitem[\protect\citeauthoryear{{Strateva}, {Ivezi{\'c}}, {Knapp},
  {Narayanan}, {Strauss}, {Gunn}, {Lupton}, {Schlegel}, {Bahcall}, {Brinkmann},
  {Brunner}, {Budav{\'a}ri}, {Csabai}, {Castander}, {Doi}, {Fukugita}, {Gy{\H
  o}ry}, {Hamabe}, {Hennessy}, {Ichikawa}, {Kunszt}, {Lamb}, {McKay},
  {Okamura}, {Racusin}, {Sekiguchi}, {Schneider}, {Shimasaku}, \&
  {York}}{{Strateva} et~al.}{2001}]{strateva01}
{Strateva} I., et al.\ 2001, AJ, 122, 1861

\bibitem[\protect\citeauthoryear{{Tanaka}, {Kodama}, {Arimoto}, {Okamura},
  {Umetsu}, {Shimasaku}, {Tanaka} \& {Yamada}}{{Tanaka} et~al.}{2005}]{tanaka}
{Tanaka} M.,  {Kodama} T.,  {Arimoto} N.,  {Okamura} S.,  {Umetsu} K.,
  {Shimasaku} K.,  {Tanaka} I.,    {Yamada} T.,  2005, MNRAS, 362, 268

\bibitem[\protect\citeauthoryear{{Trager}, {Faber}, {Worthey} \&
  {Gonz{\'a}lez}}{{Trager} et~al.}{2000}]{trager00}
{Trager} S.~C.,  {Faber} S.~M.,  {Worthey} G.,    {Gonz{\'a}lez} J.~J.,  2000,
  AJ, 120, 165

\bibitem[\protect\citeauthoryear{{Weinmann}, {van den Bosch}, {Yang} \&
  {Mo}}{{Weinmann} et~al.}{2006{\natexlab{a}}}]{wm06}
{Weinmann} S.~M.,  {van den Bosch} F.~C.,  {Yang} X.,    {Mo} H.~J.,  2006{\natexlab{a}},
  MNRAS, 366, 2

\bibitem[\protect\citeauthoryear{{Weinmann}, {van den Bosch}, {Yang}, {Mo},
  {Croton} \& {Moore}}{{Weinmann} et~al.}{2006{\natexlab{b}}}]{weinmann06}
{Weinmann} S.~M.,  {van den Bosch} F.~C.,  {Yang} X.,  {Mo} H.~J.,  {Croton}
  D.~J.,    {Moore} B.,  2006{\natexlab{b}}, MNRAS, 372, 1161




\bibitem[\protect\citeauthoryear{{Wolf}, {Aragon-Salamanca}, {Balogh}, {Barden},
  {Bell}, {Gray}, {Peng}, {Bacon}, {Barazza}, {Boehm}, {Caldwell}, {Gallazzi},
  {Haeussler}, {Heymans}, {Jahnke}, {Jogee}, {van Kampen}, {Lane}, {McIntosh},
  {Meisenheimer}, {Papovich}, {Sanchez}, {Taylor}, {Wisotzki}, \&
  {Zheng}}{{Wolf} et~al.}{2008}]{dusty}
{Wolf} C., et al. 2009, MNRAS, 393, 1302

\bibitem[\protect\citeauthoryear{{Wolf}, {Gray} \& {Meisenheimer}}{{Wolf}
  et~al.}{2005}]{wolf05}
{Wolf} C.,  {Gray} M.~E.,    {Meisenheimer} K.,  2005, A\&A, 443, 435

\bibitem[\protect\citeauthoryear{{Worthey}}{{Worthey}}{1994}]{worthey}
{Worthey} G.,  1994, ApJS, 95, 107

\bibitem[\protect\citeauthoryear{{Yang}, {Mo}, {van den Bosch}, {Pasquali}, {Li},
  \& {Barden}}{{Yang} et~al.}{2007}]{weinmann}
{Yang} X., {Mo} H.~J., {van den Bosch} F.~C., {Pasquali} A., {Li} C., \&
  {Barden} M. 2007, ApJ, 671, 153 (Yang07)

\bibitem[\protect\citeauthoryear{{Yee} \& {Ellingson}}{{Yee} \&
  {Ellingson}}{2003}]{YE}
{Yee} H.~K.~C.,  {Ellingson} E.,  2003, ApJ, 585, 215

\bibitem[\protect\citeauthoryear{{Yee} \& {L{\'o}pez-Cruz}}{{Yee} \&
  {L{\'o}pez-Cruz}}{1999}]{bgc}
{Yee} H.~K.~C.,  {L{\'o}pez-Cruz} O.,  1999, AJ, 117, 1985

\bibitem[\protect\citeauthoryear{{York}, {Adelman} \& {Anderson et al.}}{{York}
  et~al.}{2000}]{sloan}
{York} D.~G.,  {Adelman} J.,    {Anderson et al.} 2000, AJ, 120, 1579

\end{thebibliography}
\end{document}